\documentclass{aa}   
\usepackage{graphicx}
\usepackage{txfonts}
\usepackage[colorlinks=true,allcolors=blue]{hyperref}
\usepackage{booktabs}
\usepackage{multirow}

\begin{document}

\title{ Multiwavelength study of extreme variability in LEDA 1154204: A changing-look event in a type 1.9 Seyfert}
   \subtitle{}
   \titlerunning{Flaring event in CLAGN LEDA 1154204}
   \authorrunning{T. Saha et al.}
   \author{T. Saha \inst{\ref{camk},\ref{aip}}
			\and A. Markowitz \inst{\ref{camk},\ref{ucsd}}
			\and D. Homan \inst{\ref{aip}}
			\and M. Krumpe \inst{\ref{aip}}
			\and S. Haemmerich  \inst{\ref{remeis}}
			\and B. Czerny\inst{\ref{cft}}
			\and M. Graham \inst{\ref{caltech}}
			\and S. Frederick \inst{\ref{umd},\ref{vb}}
			\and M. Gromadzki \inst{\ref{uw}} 
			\and S. Gezari \inst{\ref{stsci}}
			\and H. Winkler \inst{\ref{uj}}
			\and D. A. H. Buckley \inst{ \ref{saao}, \ref{ucpt}, \ref{ufs}}
			\and J. Brink \inst{\ref{aip},\ref{ucpt},\ref{saao}}
			\and M. H. Naddaf \inst{\ref{camk},\ref{cft}}
			\and A. Rau \inst{\ref{mpe}}
			\and J. Wilms\inst{\ref{remeis}}
			\and A. Gokus \inst{\ref{wu},\ref{remeis},\ref{wuerzburg}}
			\and Z. Liu \inst{\ref{mpe}}
			\and I. Grotova\inst{\ref{mpe}}
          }

   \institute{Nicolaus Copernicus Astronomical Center of the Polish Academy of Sciences, ul.\ Bartycka 18, 00-716 Warszawa, Poland\label{camk}\\
   email: tathagata@camk.edu.pl
			\and Center for Astrophysics and Space Sciences, University of California, San Diego, 9500 Gilman Drive, La Jolla, CA 92093-0424, USA\label{ucsd}
                \and Leibniz-Institut für Astrophysik Potsdam, An der Sternwarte 16, 14482 Potsdam, Germany \label{aip}
                \and Dr. Karl Remeis-Observatory \& ECAP, Friedrich-Alexander-Universit\"at Erlangen-N\"urnberg, Sternwartstr.\ 7, 96049 Bamberg, Germany\label{remeis}
                \and Center for Theoretical Physics, Polish Academy of Sciences, Al. Lotnik\'ow 32/46, 02-668, Warszawa, Poland\label{cft}
                \and Division of Physics, Mathematics, and Astronomy, California Institute of Technology, Pasadena, CA 91125, USA \label{caltech}
                \and Department of Astronomy, University of Maryland, College Park, MD 20742, USA \label{umd}
               \and Department of Astronomy, Vanderbilt University, 6301 Stevenson Center, Nashville, TN 37235, USA \label{vb}
               \and Astronomical Observatory, University of Warsaw, Al.\ Ujazdowskie 4, 00-478 Warszawa, Poland \label{uw} 
               \and Space Telescope Science Institute, 3700 San Martin Drive, Baltimore, MD 21218, USA \label{stsci}
               \and Department of Physics, University of Johannesburg, Kingsway, 2006 Auckland Park, Johannesburg, South Africa \label{uj}
               \and South African Astronomical Observatory, PO Box 9, Observatory 7935, Cape Town, South Africa \label{saao}
               \and Department of Astronomy, University of Cape Town, Private Bag X3, Rondebosch 7701, South Africa \label{ucpt}
                \and Department of Physics, University of the Free State, PO Box 339, Bloemfontein 9300, South Africa \label{ufs}
                \and Max Planck Institute for Extraterrestrial Physics, Giessenbachstrasse, D-85741 Garching, Germany\label{mpe}
                \and Department of Physics \& McDonnell Center for the Space Sciences, Washington University in St. Louis, 1 Brookings Drive, St.~Louis, MO 63130, USA \label{wu}
               \and Lehrstuhl f\"ur Astronomie, Universit\"at W\"urzburg, Emil-Fischer-Stra{\ss}e 31, 97074 W\"urzburg, Germany\label{wuerzburg}
             }

   \date{Received; accepted}

\abstract{Multiwavelength studies of transients in actively accreting supermassive black holes have revealed that large-amplitude variability is frequently linked to significant changes in the optical spectra -- a phenomenon known as changing-look AGN (CLAGN).}
{In 2020, the Zwicky Transient Facility detected a transient flaring event in the type 1.9 AGN LEDA 1154204, wherein brightness sharply increased by 0.55 mag in one month, then began to decay. Spectrum Roentgen Gamma (SRG)/eROSITA also observed the object as part of its all-sky X-ray surveys, after the flare had started decaying.}
{We performed a three-year, multiwavelength follow-up campaign to track the source's spectral and temporal characteristics, during the post-flare fading. This campaign included optical spectroscopy, X-ray spectroscopy and photometry, and UV, optical, and IR continuum photometry.}
{Optical spectra taken near the flare peak revealed a broad double-peaked H$\beta$ emission and a blue continuum, both undetected in a 2005 archival spectrum; broad H$\beta$ had increased by a factor $>$5--6.  Then, from late 2020 through 2023, broad Balmer line flux faded as the continuum faded, with Balmer decrement increasing by $\sim$2.2, consistent with the expected ionization response. The X-ray spectrum exhibits no significant spectral variability despite dramatic flux variation -- a factor of 17. There is no evidence of a soft X-ray excess, indicating an energetically unimportant warm corona.}
{The transient event was likely triggered by a disk instability in a pre-existing AGN-like accretion flow, culminating in the observed multiwavelength variability -- X-rays via thermal Comptonization, BLR illumination, and IR dust echo -- and CLAGN event.}

\maketitle

\section{Introduction}
A supermassive black hole (SMBH) at the center of a galaxy grows through long-term accretion of matter \citep{soltan1982}, powering active galactic nuclei (AGNs). AGNs emit across multiple electromagnetic wavebands. Among the many ways to classify them is on the basis of optical spectral emission line properties: so-called type~1 sources exhibit both broad and narrow emission lines, while type~2 sources exhibit only narrow emission lines \citep[][]{khachikian1974}. Intermediate sub-types are classified depending on the relative intensities of the broad Balmer emission lines \citep[][]{osterbrock1976, winkler1992}\footnote{In this paper, we consider sub-types 1.0, 1.2, 1.5, 1.8, and 1.9 to be sub-types of type 1, as broad H$\alpha$ is present in all such cases. We follow e.g., \citet{osterbrock1981}, \citet{winkler1992} and \citet{runco2016} in defining type 1.9 as having broad H$\alpha$ detectable against the local continuum while broad H$\beta$ is not detectable, and type 1.8 in having broad H$\beta$ detectable but weak. Across types 1.0, 1.2, 1.5, 1.8, the strength of broad H$\beta$ flux relative to e.g., [\ion{O}{iii}] flux \citep{winkler1992} or narrow H$\beta$ flux \citep[e.g.,][]{osterbrock1977,osterbrock1981} gradually decreases.}. A typical AGN -- Seyfert or quasar -- exhibits stochastic multi-wavelength variability on timescales ranging from hours to decades \citep[reviews are given by e.g.,][]{mushotzky1993} where flux can vary up to a factor of $10-20$ in X-rays \cite[e.g.][]{markowitz2004} and by factors of few in the optical/UV \cite[e.g.][]{uttley2004}. One of the leading models explaining such variability trends is the existence of inward-propagating fluctuations in local mass accretion rate \citep[e.g.][]{ingram2011}, aided by local magneto-rotational instability \citep[MRI;][]{balbus1991}.

It is not clear whether accretion activity in all AGNs occurs with steady accretion rates over long timescales ($\sim$ $10^6$ to $10^7$~yr), or through episodic extreme variability in accretion rates over shorter timescales \cite[e.g.][]{shen2021}. Clues to this question have been arriving in the form of "changing-look AGN" (CLAGNs), sources wherein one or more broad Balmer emission lines (e.g., H$\alpha$, H$\beta$) can substantially strengthen or weaken -- or even appear or disappear -- on timescales from a few months to several years. In some cases, sources can switch between type 1 and type 2, as broad Balmer lines are observed to appear or disappear completely \citep[e.g.,][]{macleod2016,trakhtenbrot2019b}. In other cases, one observes changes across type~1 sub-type, driven by evolution in the strength of broad H$\beta$ relative to that of broad H$\alpha$ \citep[e.g.,][]{runco2016,green2022}. In the last couple decades, well over two hundred AGNs have been discovered to change classification to/from type 1, 2, or intermediate subtypes \citep[e.g.,][]{trippe2008, denny2014, shappee2014, lamassa2015, macleod2016, ruan19, panda2024, guo2024, yang2024}. Critically, the bulk of these optical spectral changes are tied to extreme variations in optical, UV, and/or X-ray continuum flux, with amplitude significantly higher than that measured in the case of regular stochastic variability. Such variability indicates major variations in the accretion rate, which drives variations in the ionizing flux that the broad line region (BLR) sees. Optical spectral changes driven by variations in line-of-sight obscuration associated with dusty clouds or winds \citep[e.g.][]{miniutti2014,mehdipour2017,zeltyn2022} occur only very rarely.

In Jan.--Mar.\ 2020, the Zwicky Transient Facility \citep[ZTF;][]{bellm2019} observed a strong optical flux rise over a timescale of 34~days in a type-1.9 Seyfert galaxy, LEDA 1154204. In parallel, the Extended ROentgen Survey with an Imaging Telescope Array \cite[eROSITA;][]{predehl2021}, the soft X-ray telescope onboard the \textit{Spectrum Roentgen/Gamma} (\textit{SRG}) spacecraft \citep{sunyaev2021} observed decaying X-ray emission in LEDA 1154204 through its four successive all-sky scans spanning March 2020 to September 2021. We obtained X-ray monitoring to track the X-ray coronal emission, optical and UV photometry to track the thermal emission from the accretion disk, infrared photometry to track the thermal emission from parsec-scale circumnuclear dust and optical spectroscopy to track the behavior of the broad line region (BLR). As demonstrated below, we track how the optical spectrum changed from type 1.9  -- no broad H$\beta$ emission detected in 2005-- to type 1.0 in 2020, coincident with optical continuum flaring \citep{frederick2020atel}. Our campaign subsequently tracked concurrent fading in multi-band continuum fluxes and broad Balmer fluxes over the next three years.

The remainder of this paper is organized as follows: Sect.~\ref{sec:object_properties} discusses the detection of the transient continuum flare. In Sect.~\ref{sec:data_reduction}, we report all the follow-up data  and their reduction. Sect.~\ref{sec:analysisandresults} describes all analysis, including the multi-band continuum variability, X-ray spectroscopy, broadband SED fitting, and modeling of the optical emission line spectra. In Sect.~\ref{sec:disc}, we discuss our results in the context of variable accretion processes and the possible nature of the flare, which we conclude is the root cause behind the observed multi-band continuum variability and the changing-look transitions in the optical spectrum. We summarize our conclusions in Sect.~\ref{sec:summary}.

\section{Flare detection and counterpart}\label{sec:object_properties}
ZTF's large-area photometric monitoring, with a cadence of once every few days, enabled detection of the flaring event at coordinates of $\alpha$= 04h28m38.77s, $\delta$=$-$00d00m39.7s. The event was designated as AT2019aabw/ZTF19aagwzod \citep{frederick2020atel}. We found that (Sec. \ref{sec:optical_lc}, below) the $g$- and $r$-band reduced magnitudes decreased by $|\Delta g| \simeq 0.55$ and $|\Delta r| \simeq 0.29$ in 34~days, from 2020 Jan.\ 5 to Feb.\ 8. In parallel, eROSITA's all-sky X-ray scans (eRASS), tracked the decay in X-ray flux in a point source located at $\alpha $= 04h28m38.58s and $\delta=$ $-$00d00m41.8s, with a positional uncertainty of $\sim$1$\arcsec$. 0.2--5.0 keV flux fell steadily from (3.2$\pm$0.4) $\times 10^{-12}$ erg cm$^{-2}$ s$^{-1}$ during eRASS1 (2020 Mar.\ 5--6) to (2.1$\pm$1.2) $\times 10^{-13}$ erg cm$^{-2}$ s$^{-1}$ during eRASS4 (2021 Aug.\ 28--29). The event is designated eRASSt~J042838.58$-$000041.81, henceforth referred to as J0428$-$00 for simplicity.

For both optical and X-ray detections, the most likely counterpart is LEDA 1154204, located at $\alpha =$ 04:28:38.77, $\delta$=$-$00:00:40\footnote{\url{https://ned.ipac.caltech.edu/}}.
An optical spectrum was taken in 2005 (target ID g0428388-000040) as part of the 6dFGRS survey \citep{jones2005}; \citet{jones2009} established its redshift to be 0.070.

\section{Data acquisition and reduction}\label{sec:data_reduction}
\begin{table*}
\centering
\caption{X-ray and space-based optical/UV observations of J0428$-$00.}\label{tab:X-ray_UV_observations}
\begin{tabular}{ccccccl}

 \hline
 Instrument & ObsID       & Date (MJD) & X-ray Exp. (in ks)  & Optical \& UV filters (exp. in ks)\\
 \hline
\textit{Swift}          & 00013199001  & 58886     & 2.0     & V(0.17), B(0.17), U(0.17),  W1(0.3), M2(0.37), W2(0.67)&  \\
\textit{Swift}          & 00013199003  & 58895     & 1.4     & V(0.1), B(0.1), U(0.1),  W1(0.20), M2(0.35), W2(0.44)&  \\                     
\textit{Swift}          & 00013199004  & 58899     & 0.7      & B(0.09),  U(0.09),  W1(0.26), W2(0.2) &      \\
\textit{Swift}          & 00013199005  & 58901     & 2.1     & V(0.17), B(0.17), U(0.17),  W1(0.3), M2(0.5), W2(0.7)& \\
\textit{Swift}          & 00013199006  & 58907     & 1.9    & V(0.15), B(0.15), U(0.15),  W1(0.3), M2(0.47), W2(0.61)&  \\
eROSITA        &   eRASS1    & 58911     & 0.3 (0.14)      & NONE \\
\textit{Swift}          & 00013199007  & 58912     & 1.2    & V(0.11), B(0.11), U(0.11), W1(0.21), M2(0.22), W2(0.42)& \\
\textit{Swift}          & 00013199008  & 58914     & 1.7     & V(0.13), B(0.13), U(0.13), W1(0.26), M2(0.42) ,W2(0.53)& \\
\textit{Swift}          & 00013199009  & 58916     & 1.7     & V(0.14), B(0.14), U(0.14), W1(0.28), M2(0.42), W2(0.56)& \\
\textit{Swift}          & 00013199010  & 58919     & 1.7     & V(0.14), B(0.14), U(0.14), W1(0.28), M2(0.42), W2(0.55) & \\
eROSITA        &   eRASS2    & 59093     & 0.3 (0.14)      & NONE \\
eROSITA        &   eRASS3    & 59263     & 0.3 (0.14)     & NONE \\
eROSITA        &   eRASS4   & 59452     & 0.3 (0.14)     & NONE \\
\textit{Swift}          & 00014932001  & 59550     & 5.7     & V(0.29), B(0.29), U(0.29), W1(0.85), M2(2.3), W2(1.4) & \\
\textit{XMM-Newton}     & 0903991001   & 59651     & 21.0   & V(3.5),B(3.5), W1(4.4), M2(4.4), W2(4.4)     \\
\textit{Swift}          & 00015274001  & 59786    & 5.5     & V(0.43), B(0.43), U(0.43), W1(0.87), M2(1.44), W2(1.74) & \\
\textit{XMM-Newton}     & 0903991101   & 60017 & 40.0 &  V(4.4), B(4.4), U(4.4), M2(4.4)\\
\hline
\end{tabular}
\tablefoot{Log of space-based X-ray, optical, and ultraviolet observations with eROSITA, \textit{Swift} (XRT and UVOT instruments), and \textit{XMM-Newton} (EPIC-pn and OM). The exposure times listed in the table for \textit{Swift} and \textit{XMM-Newton} are good time intervals (GTI) after screening. For the eROSITA observations we report both the GTI and the vignetting-corrected exposure (in parentheses). In this table, W1, M2 and, W2 are abbreviations for the UVW1, UVM2, and UVW2 filters, respectively.}
\end{table*}

\begin{figure*}
\centering
\includegraphics[scale=0.39]{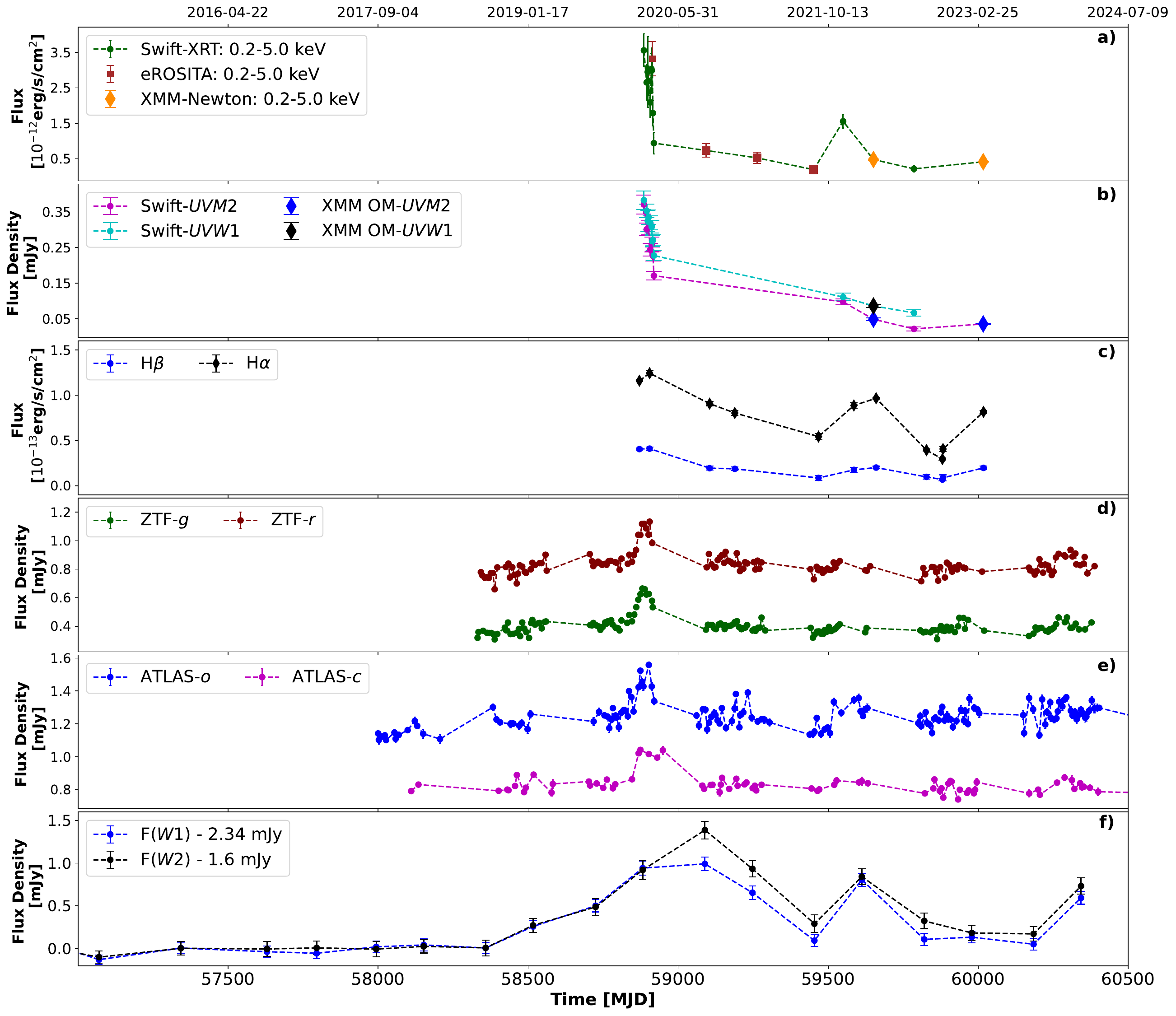}
\caption{Optical, UV, and X-ray light curves for the entire period of our monitoring. From top to bottom are plotted: a) 0.2--5.0 keV flux from \textit{Swift}-XRT, \textit{XMM-Newton}-EPIC, and eROSITA,
b) UVM2- and UVW1-band flux densities from \textit{Swift}-UVOT and \textit{XMM-Newton}-OM, and
c) line fluxes for broad H$\beta$ and H$\alpha$ emission components, integrated over the whole broad profile (Sect.~\ref{subsec:2-gauss}).
d) ZTF $g$- and $r$-band, 
e) ATLAS $c$- and $o$-band,
f) \textit{WISE/NEOWISE} W1 and W2 bands
In panels a), b), d), and e), there are some data points for which the error bars are smaller than the data point marker.
}\label{fig:all_lc}
\end{figure*}

\subsection{Optical and IR photometric data}
We obtained publicly-available $g$- and $r$-band photometric points taken with ZTF from the Infrared Science Archive (IRSA) website\footnote{\url{https://irsa.ipac.caltech.edu/Missions/ztf.html}}. We used data from Data Release 22; the online light curve generation tool provides all measurements for a given source (here, ZTF catalog ID number 405213300018639) from the calibrated single-exposure PSF-fit-derived catalogs \citep{masci2018}. We converted AB magnitudes to mJy using standard zeropoints. 

We ran the online forced-photometry pipeline of ATLAS \footnote{\url{https://fallingstar-data.com/forcedphot/}} \citep{tonry2018, smith2020, shingles2021} to generate the $o$- (560--820 nm) and $c$- (420--650~nm) band optical light curves.  This pipeline calculates a point-spread-function for each image based on high signal/noise stars in the field view; we use reduced-image, not difference-image, data. 

We obtained public IR photometric monitoring courtesy of the \textit{Wide-field Infrared Survey Explorer} (\textit{WISE/NEOWISE}) \citep{wright2010, mainzer2014}. We downloaded all available single-epoch W1 (3.4~$\mu$m) and W2 (4.6~$\mu$m) band PSF-fitted
photometry on J0428$-$00 from the AllWISE Multiepoch Photometry Table (MJD range 55244.8 to 55440.9)
and the NEOWISE-R Single Exposure Source Table (MJD range 56709.3 to 60343.2, the most recent data released as of this writing) from IRSA\footnote{\url{https://irsa.ipac.caltech.edu/Missions/wise.html}}.
The \textit{WISE} zero magnitude attributes are reported in
\cite{jarrett2011}; we converted the Vega magnitudes to mJy, and then rebinned the light curves to one point every six months. 

The resulting  ZTF, ATLAS, and \textit{WISE/NEOWISE} light curves are plotted in Fig.~\ref{fig:all_lc}.

\subsection{SRG/eROSITA}
eROSITA detected the object during each of its first four all-sky scans, eRASS1--4. 
We combined data from all seven Telescope Modules.
We used event data from processing version c020, and the eROSITA data analysis software eSASS version \textit{eSASSuser 211214} \citep{brunner2022} in 
High Energy Astrophysics Software (HEASOFT) version 6.29. For each dataset, a circle and an annulus were used for the source and background regions, respectively. The radii were selected based on source brightness, where larger regions correspond to the source being brighter. Source region sizes varied between 105$\arcsec$ (most bright) and 47$\arcsec$ (least bright). Neighboring sources detected were excluded from the background regions in order to avoid background contamination. We list the eROSITA observations in Table~\ref{tab:X-ray_UV_observations}.

\subsection{Swift}
We observed J0428$-$00 twelve times using the \textit{Neil Gehrels Swift Observatory} \citep[\textit{Swift;}][]{gehrels2004}. Ten pointings occurred during the initial flaring state between Feb.\ and Mar.\ 2020; the 11th observation occurred in Dec.\ 2021; the final observation occurred in Mar.\ 2023. 

The X-ray Telescope (XRT) exposures varied from 0.5 to 5.7 ks (Table~\ref{tab:X-ray_UV_observations}). We calibrated the event files using \texttt{xrtpipeline} in HEASOFT version 6.29 and the latest calibration files. All spectra were extracted with circular regions of 40$\arcsec$ for both the source and the background. We generated the ancillary response files using \texttt{xrtmkarf}, and the response matrix (RMF) was taken from the latest calibration database.

For the Ultraviolet/Optical Telescope (UVOT) data, we selected a circular region of 5$\arcsec$ for the source and a 25$\arcsec$ circular region located a few arcminutes from the source. for the background for images taken by each of the UVOT filters. We used the task \texttt{uvotsource} to perform photometry and estimate fluxes for each of the available filters for the given observation. 
We applied Galactic extinction correction externally following the extinction estimates from \cite{schlafly2011}.
We also used the task \texttt{uvot2pha} to generate \textsc{Xspec}-readable spectral files for the purpose of modeling simultaneous optical/UV/X-ray SEDs.

\subsection{\textit{XMM-Newton}}
Two \textit{XMM-Newton} \citep{jansen2001} observations (PI: M.\ Krumpe) with total durations of 38 ks and 55 ks were taken on 23 Mar.\ 2022 and 14 Mar.\ 2023 respectively (Table~\ref{tab:X-ray_UV_observations}). The EPIC observations were taken in full window mode. We reduced the EPIC-pn data with SAS  v21.0.0 and HEASOFT v6.28 using standard procedures for point sources. A 40$\arcsec$  circular region centered around the source was used to extract the source spectrum. The background spectrum was extracted from a circular source-free region of the same radius, a few arcmin away from the source. After high-background cleaning 21.0~ks and 40.0~ks of good EPIC-pn integration time is left, corresponding to 0.2--10 keV spectral counts of $3.31\times 10^3$ and $5.34\times 10^3$ respectively. We did not find any evidence of pileup in either spectrum, as per the XMMSAS task \texttt{epatplot}.

We reduced the Optical Monitor (OM) imaging mode  data using the \texttt{omichain} pipeline processing. This command applies flat fielding, source detection, and aperture photometry, and ultimately creates mosaiced images.
We used the \texttt{om2pha} command to generate \textsc{Xspec}-readable spectral files for all the available filters (Table~\ref{tab:X-ray_UV_observations}). 

\subsection{Optical Spectroscopy}
\begin{table*}
    \centering
    \caption{Optical spectroscopic observations of J0428$-$00 during 2020--2023}
    \begin{tabular}{@{}ccccccc@{}}\toprule
        & Date & Date & Instrument (Telescope)   & Slit width & Seeing & Exposure \\
        & (MJD) & & & ($\arcsec$) & ($\arcsec$) & (s)\\
        \midrule
        \#1 & 58871 & 2020-01-23 & LRIS+LRISBLUE (Keck)  & 1.0 & 0.46        & 600\\
        \#2 & 58905 & 2020-02-26 & DeVeny (LDT)          & 1.5 & 1.5         & 1800\\
        \#3 & 59105 & 2020-09-13 & DeVeny (LDT)          & 1.5 & 1.5         & 1700\\
        \#4 & 59189 & 2020-12-06 & DeVeny (LDT)          & 1.5 & 1.5         & 2400\\
        \#5 & 59468 & 2021-09-11 & SpUpNIC (SAAO)        & 2.7 & $\sim$2.5   & 2400\\
        \#6 & 59586 & 2022-01-07 & RSS (SALT)            & 1.5 & 2.7   & 540 \\
        \#7 & 59660 & 2022-03-22 & FORS2 (VLT)           & 1.3 & $<$1.3        & 900\\
        \#8 & 59828 & 2022-09-06 & DBSP                  & 1.5 & 1.5         & 600  \\
        \#9 & 59881 & 2022-10-29 & RSS (SALT)            & 1.5 & 1.4     & 500 \\
        \#10 & 59883 & 2022-10-31 & RSS (SALT)            & 1.5 & 3.5--4.5 & 720  \\
        \#11 & 60018 & 2023-05-15 & FORS2 (VLT)           & 1.3 & $<$1.3        & 900 \\
    \bottomrule
    \end{tabular}
    \label{tab:opt_spectra}
\end{table*}

We obtained eleven long-slit spectra from Jan.\ 2020 to Mar.\ 2023 (referred to as spectra \#1--11), as summarized in Table~\ref{tab:opt_spectra}. We obtained one spectrum at each of the following facilities: the 10~m class Keck telescope \citep{oke1995}; the 5~m class Hale Telescope\footnote{\url{http://sites.astro.caltech.edu/palomar/about/telescopes/hale.html}}; and the 1.9~m telescope \citep{crause2019} at South African Astronomical Observatory (SAAO). 
We also obtained two spectra from the 8~m class Very Large Telescope \citep[VLT,][]{appenzeller1998}, a part of the European Southern Observatory (ESO). 
We also obtained three spectra each using the 4~m class Lowell Discovery Telescope (LDT)\footnote{\url{https://lowell.edu/research/telescopes-and-facilities/ldt/}} and the 10~m class Southern African Large Telescope \cite[SALT;][]{buckley2006}. All observations were made with the slit at parallactic angle. The spectral resolution of all our optical spectra ranged between 400 and 1300.

The Keck spectrum was reduced using the LPipe package. The LDT and SALT spectra were reduced with standard IRAF routines. The VLT spectra were reduced using the EsoReflex package. The DBSP spectrum was reduced using the DBSP\_DRP program. 

For all spectra, corrections for atmospheric absorption were made following standard procedures. For example, for the SAAO spectrum, we applied corrections in the telluric A-band (7588--7700 \AA) and B-band (6855--6965 \AA). A correction for water vapor in the wavelength range 7150--7400~$\AA$ was also applied, though  the impact was negligible. 
We used the spectral atmospheric transmission coefficients from the SMARTS2 atmospheric transmission model \citep[][and references therein]{gueymard2019}. Application to the spectrum removes the telluric A and B band absorption quite adequately and the continuum looks smooth in the telluric absorption bands.

We performed flux scaling for each spectrum by fitting the [\ion{O}{III}]$\lambda$5007 line profiles following the well-established framework of \cite{groningen1992}. The method estimates a flux correction factor for the [\ion{O}{iii}]$\lambda$5007 line. It additionally corrects for wavelength-related calibration issues by cross-matching  the central wavelength of the [\ion{O}{iii}] emission line (see Appendix \ref{sec:oiii-scaling} for details).
A de-reddening correction was also applied to all optical spectra before the spectral fitting using the python package \texttt{extinction}\footnote{\url{https://extinction.readthedocs.io/en/}}, where the value of $E(B-V)$ was 0.064, obtained via the python package \texttt{sfdmap}\footnote{\url{https://github.com/kbarbary/sfdmap}} \citep[][]{schlafly2011}.

\section{Analysis and Results}  \label{sec:analysisandresults}
\subsection{Optical Continuum Variability}\label{sec:optical_lc}
All optical bands start to increase in flux dramatically starting around MJD 58855, and increase for 30--35 days before reaching maximum flux around MJD 58885--58890. Factors of increase are 1.6, 1.3, and 1.2 for ZTF $g$, ZTF $r$ and ATLAS $o$ bands, respectively (ATLAS $c$ band is relatively sparsely sampled), but these values do not account for host galaxy contamination. Optical fluxes then decay; we tracked the decrease for an additional 30 days before the sun gap, which started around MJD 58920 for $g$, $r$, and $o$ bands. Factors of decrease are 1.4, 1.15, and 1.15 for $g$, $r$, and $o$, respectively, consistent with a time-symmetric flare.
  
We searched for any interband lags via cross correlations using the Interpolated Correlation Function \citep[ICF;][]{gaskell1986,peterson1998}. All four ATLAS and ZTF light curves are well correlated with each other (maximum correlation coefficient $r_{\rm corr}$ is in the range 0.60--0.86), but there is no robust evidence for any lags or leads, with upper limits ranging from 4 to 26 days. 

There is a broad flare in both \textit{WISE} light curves, with magnitudes of increase in 0.4 (W1) and 0.6 (W2), peaking 0.5--1 year after the optical flare.
The upper limit on any \textit{WISE/NEOWISE} W1 to W2 lag is 375 days. While ICFs comparing all four optical bands to W1 yielded only tentative lags or upper limits, all four optical light curves are found to lead the W2 band, with an average lag of 174 $\pm$ 114 d. 
A variable signal from the the inner accretion flow (disk or corona) can yield a delayed reverberation signature from an extended dust structure in the IR bands. This delay would depend on the dust-sublimation radius of the system as well as on dust morphology relative to our line of sight. Assuming an average bolometric luminosity of $4\times10^{43}$~erg~s$^{-1}$, an average dust-temperature of 1500 K and, an isotropic dust distribution, using the estimate of dust submimation radius ($R_{\rm sub}$) from  \citealt{netzer2015} we obtain $R_{\rm sub} = 0.1$~parsec. This estimate corresponds to a light travel time of 120~days, consistent with the delay estimates of the W1/W2 light-curves with respect to the optical bands.

\subsection{X-ray and ultraviolet continuum}\label{sec:xray-analysis}

\begin{figure*}
\centering
\includegraphics[scale=0.32]{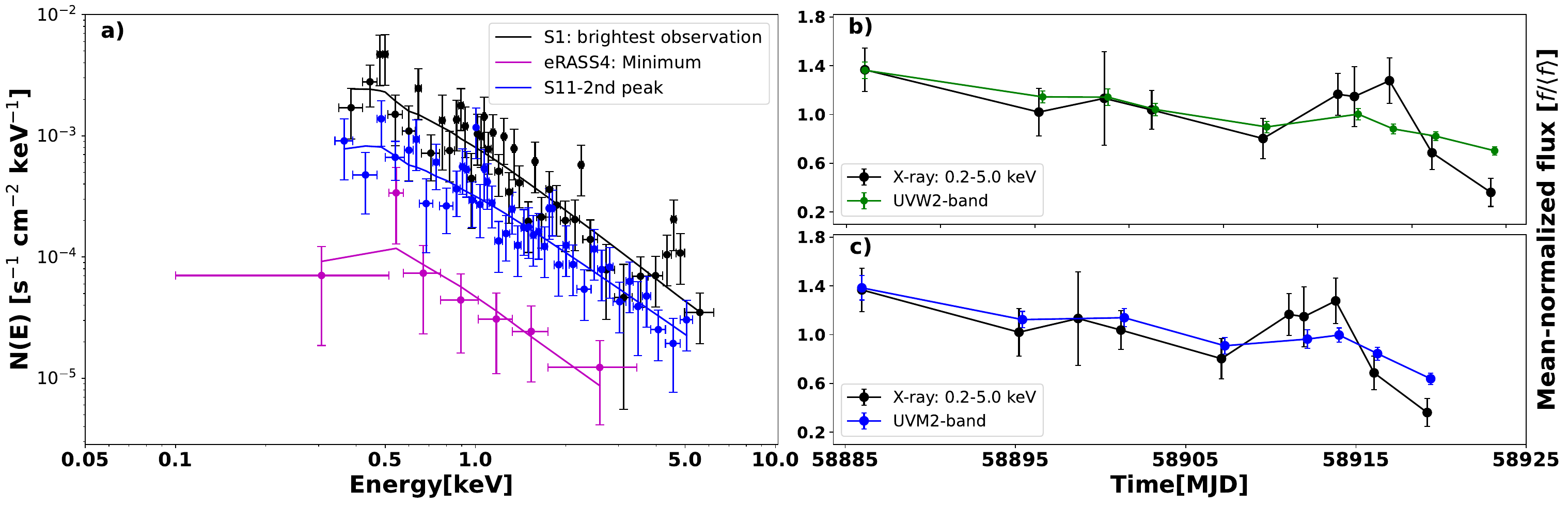}
\caption{X-ray spectra and light curves. 
Panel (a) shows the variation across three selected X-ray spectra taken during the brightest phase, the minimum flux, and during the second peak. There is no significant variation in spectral shape over time. 
Panels (b) and (c) show the short term variability in the decaying part of the flare, between MJD~58880 to 58920.  The mean-normalized light curves for 0.2--5.0~keV X-rays and the UVM2 and UVW2 bands are plotted in panels (b) and (c), respectively. 
Both the X-ray and UV light curves exhibit short term variability. 
Furthermore, both X-ray and UV light curves show overall decreases in flux by roughly the same amount, though the X-rays decrease slightly faster than the UV by the end of the campaign.}\label{fig:x-ray_spec_short_term_var}
\end{figure*}

\begin{table}
\centering
\caption{X-ray photon indices and model flux from power-law fits to all X-ray spectra.}\label{tab:x-ray}
\begin{tabular}{lccc}
\hline
Date  & Telescope  & $\Gamma_{\rm X}$ & $F_{\rm 0.2-5.0~keV}$ $^a$     \\

 (MJD)&             &         &   \\
\hline
58886 & \textit{Swift}-XRT  & $ 1.90  \pm 0.21$   & $ 3.57 \pm 0.48 $    \\
58895 & \textit{Swift}-XRT  & $ 1.75 \pm 0.36$   & $ 2.66 \pm 0.48 $   \\
58899 & \textit{Swift}-XRT  & $ 1.91 \pm 0.63$   & $ 2.98 \pm 1.00 $   \\
58901 & \textit{Swift}-XRT  & $ 1.69 \pm 0.24$   & $ 2.69 \pm 0.41 $    \\
58907 & \textit{Swift}-XRT  & $ 1.8 \pm 0.2 $   & $ 2.08 \pm 0.42 $  \\ 
58911 & eROSITA    & $ 1.73 \pm 0.34$   & $ 3.06 \pm 0.42 $  \\
58912 & \textit{Swift}-XRT  & $ 1.99 \pm 0.33$   & $ 3.0 \pm 0.6  $  \\
58914 & \textit{Swift}-XRT  & $ 1.90  \pm 0.25$   & $ 3.34 \pm 0.42 $   \\ 
58916 & \textit{Swift}-XRT  & $ 1.82 \pm 0.41$   & $ 1.79 \pm 0.38 $  \\
58919 & \textit{Swift}-XRT  & $ 1.89 \pm 0.69$   & $ 0.96 \pm 0.28 $  \\
59093 & eROSITA    & $ 2.1 \pm 0.4 $   & $ 0.74 \pm 0.21 $ \\
59263 & eROSITA    & $ 2.35 \pm 0.55$   & $ 0.52 \pm 0.15 $ \\
59452 & eROSITA    & $ 2.61 \pm 1.49$   & $ 0.19 \pm 0.12 $ \\
59550 & \textit{Swift}-XRT  & $ 1.72 \pm 0.19$   & $ 1.54 \pm 0.18 $  \\
59651 & \textit{XMM-Newton} & 1.82 $\pm$ 0.03   & $ 0.48 \pm 0.02 $ \\
59786 & \textit{Swift}-XRT  & $ 1.56 \pm 0.71$   & $ 0.22 \pm 0.09 $   \\ 
60017 & \textit{XMM-Newton} & 1.76 $\pm$ 0.02    & $0.41 \pm 0.01$ \\
\hline                          
\end{tabular}
\tablefoot{Photon indices and values of 0.2--5.0~keV absorbed flux from the power-law fits to all X-ray spectra. All errors correspond to the 90\% confidence limit of the BXA-Ultranest posteriors. The photon indices are mostly moderately flat, with $\Gamma_{\rm X}$ usually near 1.9.\\
\tablefoottext{a}{in units of $10^{-12}$ erg s$^{-1}$ cm$^{-2}$}}
\end{table}

\begin{figure*}
\centering
\includegraphics[scale=0.32]{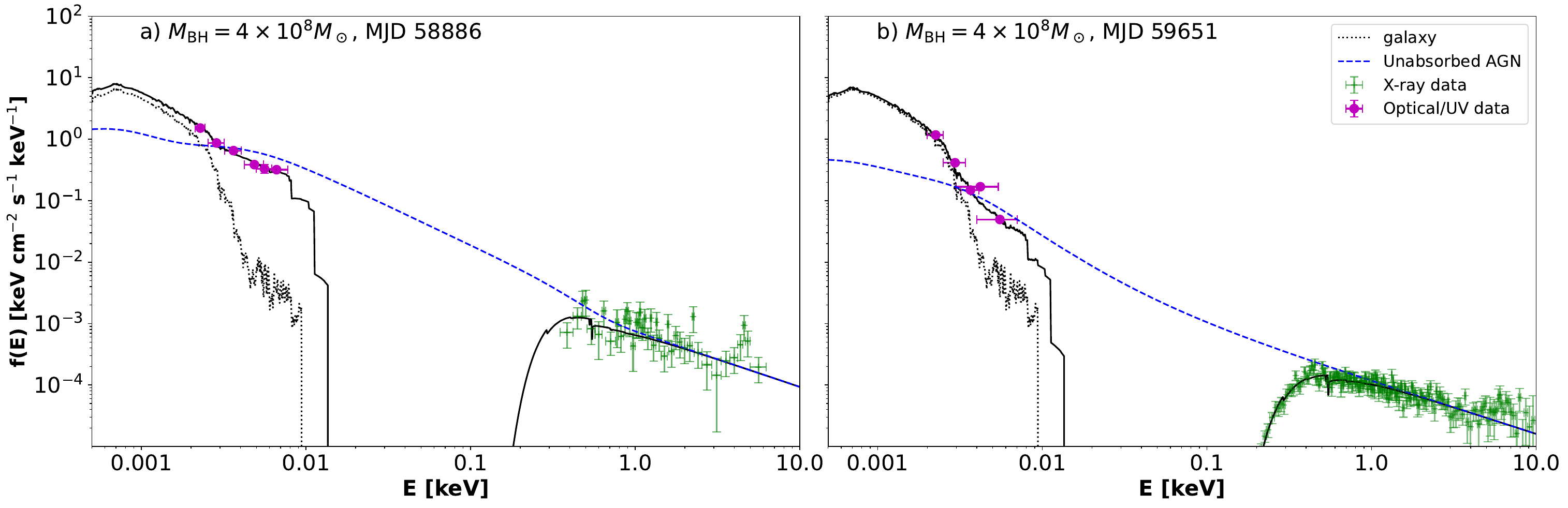}
\caption{Broadband SED fits for (a) the first \textit{Swift} SED, taken near the flare peak in Feb.\ 2020, and (b) the first \textit{XMM-Newton} SED, taken in Feb.\ 2022, assuming a black hole mass of 
$M_{\rm BH} = 4 \times 10^8\,M_{\rm \odot}$. In each plot, the blue dashed line denotes the intrinsic, unabsorbed AGN continuum.
The black dotted line denotes the host galaxy stellar contribution. The solid black line denotes the observed, Galactic-absorbed, total (AGN + host galaxy) model. In both cases, there is no significant soft X-ray excess appearing in the X-ray spectra (Sect.~\ref{sec:broadband_analysis}). In each figure, there are some data points for which the y-errorbars are smaller than the data point marker. 
}
\label{fig:broad_band}
\end{figure*}

We used the Bayesian X-ray Analysis\footnote{\url{https://johannesbuchner.github.io/BXA/index.html}} package \citep[BXA, version 4.0.6,][]{buchner2014}, implementing the nested sampling algorithm Ultranest \cite[version 3.5.5][]{buchner2021} in \textsc{xspec} (version 12.12.0) to fit the X-ray data. The fit statistic was set to \texttt{cstat}. We freeze the Galactic absorption $N_{\rm H}$ to $5.77 \times 10^{20}$~cm$^{-2}$ \citep{willingale13}. 

We fit a simple power-law model to all spectra, and acceptable fits were obtained in all cases. We find that $\Gamma_{\rm X}$ remains at relatively moderate values, with the average of best-fitting values being 1.9 (Table~\ref{tab:x-ray}) over the entire campaign, with no detected systematic variability (the scatter in best-fitting values is 0.25, compared to the average of errors being 0.4). We also tested for the presence of a soft X-ray excess \cite[e.g.][]{turner1989}, by adding a phenomenological blackbody component \textsc{zbbody}. However, no improvement was found for any spectrum\footnote{Values of Bayes factor $BF$ were less than 1; $BF \equiv Z_2/Z_1$, where $Z_1$ and $Z_2$ denote the Bayesian evidence values for the simple power-law model and the alternate model, respectively.}. The 0.5--1.0~keV \textsc{zbbody} flux was $\lesssim 10^{-4}$ of the total flux. We also tested for the presence of line-of-sight neutral obscuring gas, modeled with \textsc{zTBabs}, but no improvement to fits were obtained, with upper limits to column density around $10^{20}$~cm$^{-2}$.
We thus conclude that the X-ray spectrum can be explained by a single un-absorbed power law (Fig. \ref{fig:x-ray_spec_short_term_var}a) at all epochs.

Overall, the 0.2--5.0 keV X-ray continuum flux decays by a factor of 17 over two years with minor short-term variability superimposed on the decaying trend, as shown in (Fig.~\ref{fig:x-ray_spec_short_term_var}b). All UV bands concurrently decay by roughly the same factor, and the X-ray and UV bands are all well correlated with each other at zero lag: values of ICF correlation coefficients $r_{\rm corr}$ at zero lag are 0.884 (X-ray--UVW2), 0.915 (X-ray--UVM2), and 0.987 (UVW2--UVM2). The data sampling precludes any meaningful search for interband lags, however. Values of fractional variability amplitude $F_{\rm var}$ \citep{vaughan2003} for the X-ray, UVW2, and UVM2 light curves are $64.3 \pm 5.9\%$, $40.5\pm1.6\%$, and $50.0\pm2.3\%$, respectively.

\begin{table*}
\centering
\caption{Parameters from broad SED fits assuming a black hole mass of $4 \times 10^8 M_{\odot}$.}\label{tab:broad_band_sed_M8}
\begin{tabular}{cccccccc}
      \hline
      Date (MJD) & Telescope & $\log \lambda_{\rm Edd}$ & $\Gamma_{\rm warm}$ & $R_{\rm hot}$ ($R_{\rm g}$)& $R_{\rm warm}$ ($R_{\rm g}$)  & $\chi^2/dof$\\
      \hline
 58886     & \textit{Swift} & $-2.13 \pm 0.04$      & $2.35^{+0.12}_{-0.10}$ & $23.0 \pm 3.0$ & $300^{+170}_{-110}$   &  1.22\\
 58895     & \textit{Swift} & $-2.16 \pm 0.05$      & $2.32^{+0.16}_{-0.11}$ & $24.3^{+4.7}_{-4.1}$& $250^{+210}_{-100}$    & 0.87\\
 58899     & \textit{Swift} & $-2.23^{+0.11}_{-0.09}$      & $2.43^{+0.34}_{-0.26}$ & $22.2^{+6.1}_{-7.4}$ & $295\pm180$   &  1.06\\
 58901     & \textit{Swift}  & $-2.17 \pm 0.04$     & $2.40^{+0.16}_{-0.10}$ & $28.1 \pm 4.0$& $300 \pm 180$   & 1.34\\
 58907     & \textit{Swift}  & $-2.29^{+0.08}_{-0.05}$     & $2.60^{+0.29}_{-0.27}$ & $27.6^{+4.5}_{-6.1}$ & $209^{+256}_{-102}$   &  0.79\\
 58912     & \textit{Swift}  & $-2.26 \pm 0.05$     & $2.45^{+0.21}_{-0.16}$ & $28.6^{+7.8}_{-6.4}$& $255^{+215}_{-114}$   & 1.25\\
 58914     & \textit{Swift}  & $-2.22^{+0.06}_{-0.05}$    & $2.44^{+0.23}_{-0.21}$ & $33.7^{+7.5}_{-8.9}$& $234^{+230}_{-104}$   & 0.77\\
 58916     & \textit{Swift}  & $-2.40^{+0.10}_{-0.06}$    & $2.75^{+0.33}_{-0.36}$ & $25.2^{+4.9}_{-6.3}$& $186^{+280}_{-100}$   &  0.96 \\
 58919     & \textit{Swift}  & $-2.61^{+0.13}_{-0.07}$     & $2.97^{+0.46}_{-0.53}$ & $17.7 \pm 4.5$ & $134^{+302}_{-70}$   & 1.76\\
 59550     & \textit{Swift}  & $-2.45^{+0.04}_{-0.05}$     & $2.40^{+0.44}_{-0.21}$ & $34.0^{+7.4}_{-6.1}$& $82^{+16}_{-10}$   & 1.55\\
 59651     & \textit{XMM-Newton} & $-2.98^{+0.04}_{-0.02}$ & $2.98^{+0.67}_{-0.36}$ & $30.0 \pm 2.7$ & $149^{+264}_{-77}$  & 1.05 \\
 59786     & \textit{Swift}  & $-2.87^{+0.07}_{-0.12}$    & $2.39^{+1.71}_{-0.37}$ & $25.6^{+8}_{-5}$& $41 \pm 9$  & 6.18\\
 60017     & \textit{XMM-Newton} & $-2.99 \pm 0.01$ & $2.20^{+1.16}_{-0.17}$  & $25.9^{+2.63}_{-1.26}$& $28.7 \pm 3.0$  & 4.46\\
      \hline
\end{tabular}
\end{table*}

\subsection{Broadband SED modeling}\label{sec:broadband_analysis}

All \textit{Swift} and \textit{XMM-Newton} observations in our campaign were accompanied by simultaneous optical and/or UV photometry. We thus perform broad band spectral fitting of all datasets. The broad band spectral model contains a host galaxy template obtained from \cite{mannucci01} and the AGN broad band spectral model is \texttt{agnsed} \citep{kubota2018}.

Our model in \textsc{Xspec} notation is \texttt{redden*tbabs*(galaxy + agnsed)}. A comoving distance of $297$~Mpc \citep{wright2006}, $E(B-V) = 0.064$\footnote{\url{https://irsa.ipac.caltech.edu/applications/DUST/}} \citep{schlafly2011}, and $N_{\rm H,Gal} = 5.77 \times 10^{20}$~cm$^{-2}$ \citep{willingale13} were assumed. 
We used a black hole mass of $M_{\rm BH} = 4\times 10^8\,M_{\rm \odot}$ (see Section \ref{disc:blackhole_mass}). For all cases, we froze the value of the hard X-ray photon index ($\Gamma_{\rm hot}$ or $\Gamma_{\rm X}$) to the value obtained from the corresponding X-ray analysis. We kept the temperature of the hot corona ($k_{\rm B}T_{\rm hot}$), the temperature of the warm corona ($k_{\rm B}T_{\rm warm} $), and corona height frozen to 100~keV, 0.1~keV, and $10 R_{\rm g}$ respectively. The normalization of the host galaxy was kept frozen at the value obtained from the 2022 \textit{XMM-Newton} observation.
For all broad band analyses, we perform GW-MCMC \citep{goodman2010} fitting with a chain length of 10000, and 20 walkers. 

The values of the best-fitting parameters from our MCMC fitting and the errors corresponding to the 90\% confidence range are reported in Table \ref{tab:broad_band_sed_M8}.
We obtain acceptable fits, which indicate that in the context of the \texttt{agnsed} model, hot-Comptonization of seed photons from the accretion disk can adequately explain the X-ray spectrum of the source (Fig.~\ref{fig:broad_band}) at all stages, with no requirement for significant soft-excess emission from a warm corona. The accretion rate relative to Eddington, $\lambda_{\rm Edd} \equiv L_{\rm Bol}/L_{\rm Edd}$, decreases by a factor of 8 across our observations, from log($\lambda_{\rm Edd}$) $\simeq -2.1$ in Feb.\ 2020 to $\simeq -3.0$ in Mar.\ 2023.

\subsection{Optical spectroscopic analysis}
\begin{figure*}
\centering
\includegraphics[scale=0.55]{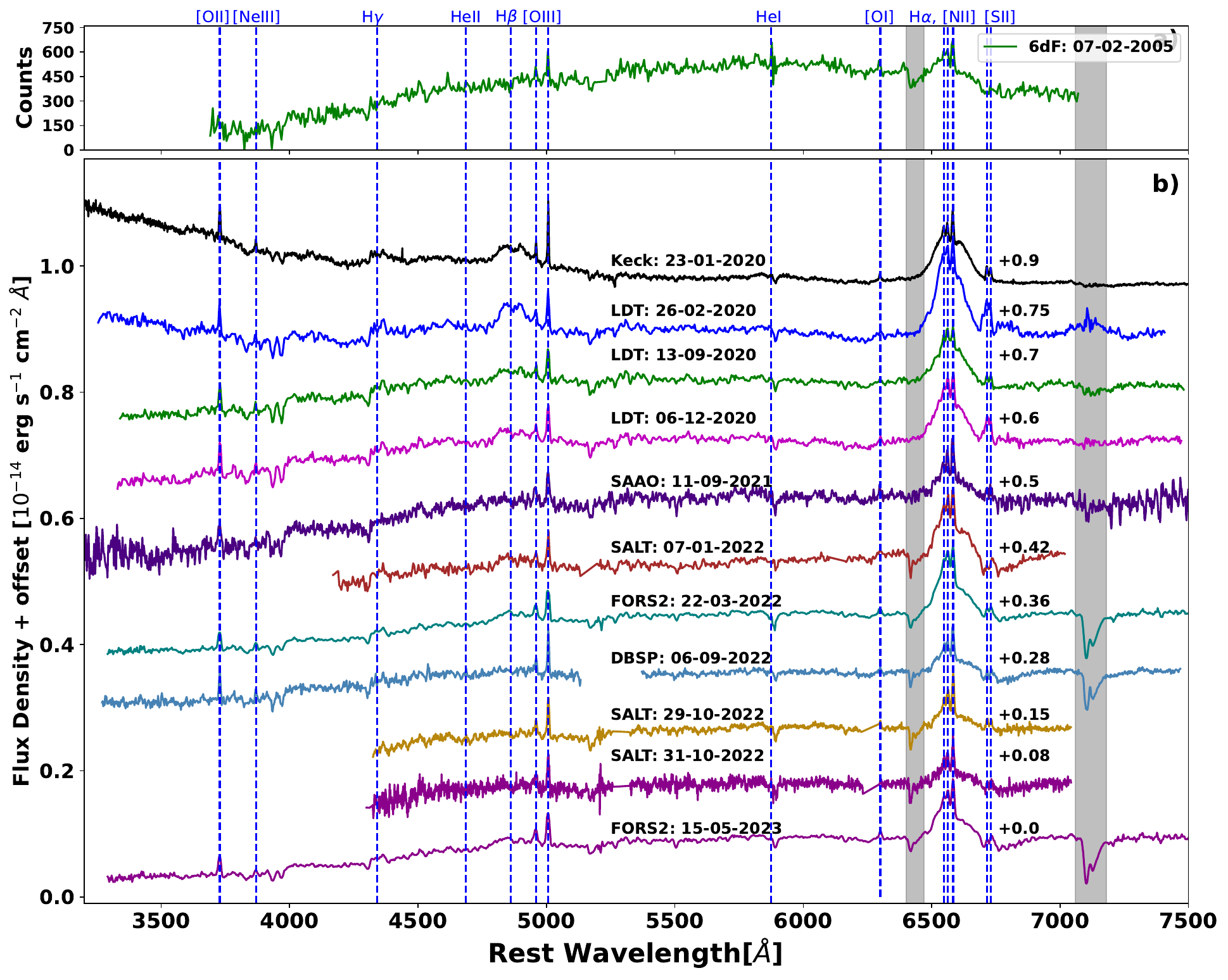} 
\caption{Optical spectra of J0428$-$00. Panel a) displays the 6dF spectrum from 2005 (not flux-calibrated), wherein broad H$\beta$ emission is not detected above the continuum. Panel b) shows the spectra taken during our 2020--2023 campaign. Spectra are not extinction corrected. 
One can follow the evolution of the broad, double-peaked H$\beta$ line, from not being detected in the 6dF spectrum, to appearing quite strongly in the 2020 spectra, but fading somewhat by 2021--2023.   
The broad H$\alpha$ profile is also double-peaked, and also fades somewhat from 2020 to 2023. 
The source also exhibits a strong blue AGN continuum in early 2020 (Keck and first LDT spectra), though it fades by the end of 2020.
The grey bands mark telluric absorption bands.}\label{fig:all_spec}
\end{figure*}

The evolution of the optical spectrum across our three-year campaign, along with the archival 6dF spectrum from 2005, is plotted in Fig.~\ref{fig:all_spec}.

\begin{figure*}
\centering
\includegraphics[scale=0.43]{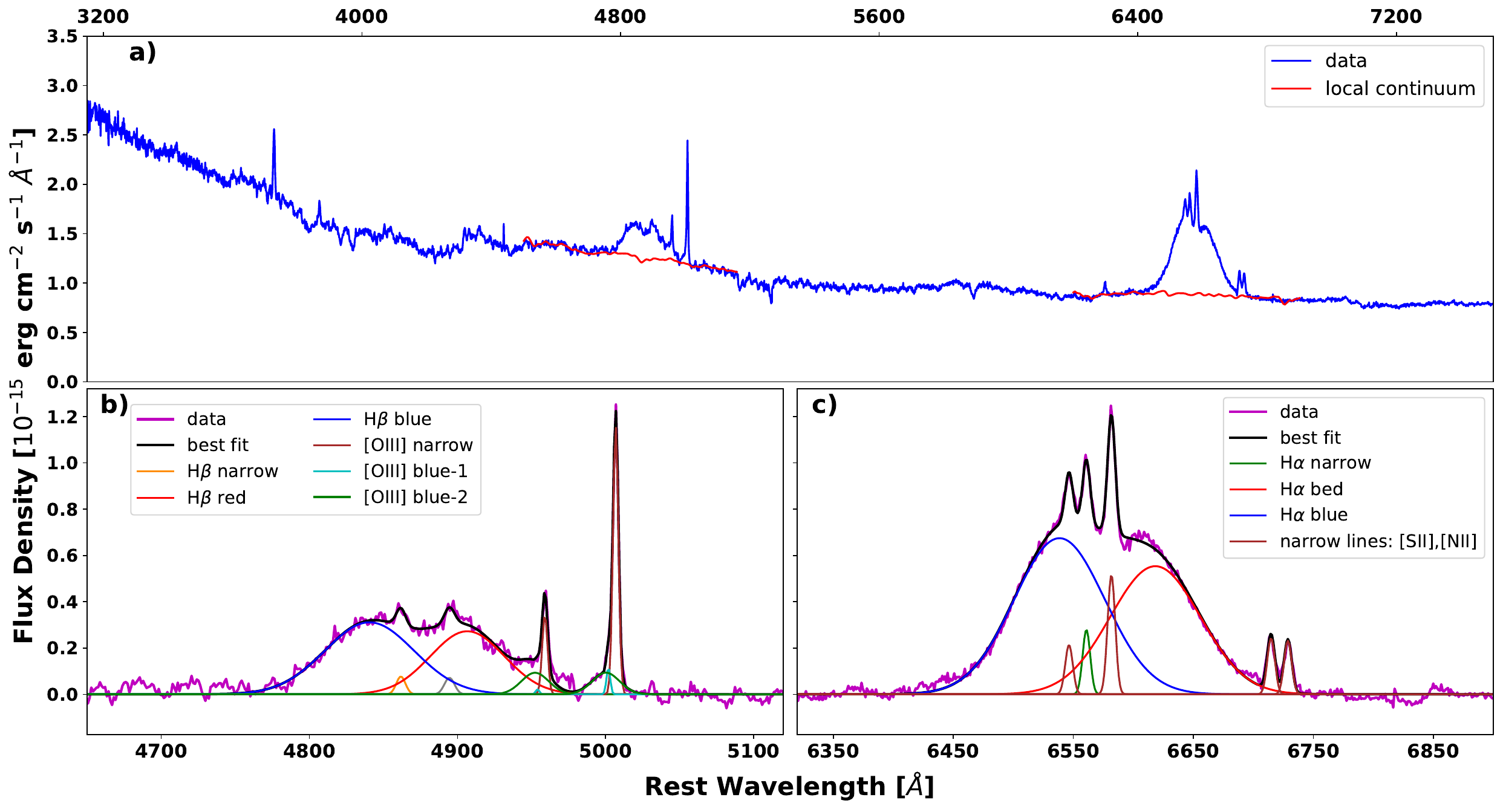}
\caption{Spectroscopic analysis of the extinction-corrected Keck spectrum. Panel (a) displays the local continuum fit. We fit the local continuum around the broad H$\beta$ and broad H$\alpha$ line regions using a power law + host-galaxy template.
Panel (b) shows the local continuum-subtracted H$\beta$ line region. The best-fitting model consists of red- and blue-shifted broad Gaussians to fit the broad H$\beta$ emission profile, as well as
Gaussian profiles fit to the narrow [\ion{O}{iii}]$\lambda \lambda$4959,5007 doublet, including two blue-shifted components for each line. The narrow grey Gaussian profile coincides with [\ion{Fe}{VII}]$\lambda$4893 emission, but such an identification is tentative (see text for details).
Panel (c) shows the local continuum-subtracted H$\alpha$ line region. The best-fitting model includes red- and blue-shifted broad Gaussians to fit the broad H$\alpha$ profile, as with H$\beta$.
There are also narrow 
Gaussian profiles fit to the [\ion{N}{ii}]$\lambda\lambda$6548,6584 and [\ion{S}{ii}]$\lambda\lambda$6716,6731 doublets, as well as to narrow H$\alpha$.}\label{fig:keck_spec}
\end{figure*}

\subsubsection{Phenomenological emission line fitting} \label{subsec:2-gauss}
J0428$-$00 exhibits the narrow emission lines typically seen in Seyfert AGNs: the [\ion{O}{iii}]$\lambda\lambda$4959,5007 doublet, the [\ion{N}{ii}]$\lambda\lambda$6548,6584 doublet, and the [\ion{S}{ii}]$\lambda\lambda$6716,6731 doublet (Fig.~\ref{fig:all_spec}). 
Additionally, the source persistently exhibits a broad H$\alpha$ emission line; as discussed below, broad H$\beta$ is detected in all spectra except the archival 6dF spectrum. Our spectral modeling is described as follows:

\begin{itemize}
\item Local continuum: For all spectra, we fit the line region over a limited wavelength range, applying a local continuum to account for the continuum flux surrounding the wings of the emission line, following standard practices \citep[e.g.][]{peterson1991}. The local continuum for the Keck (\#1) and all LDT spectra (\#2--4) is a host galaxy template plus a power law to model the AGN continuum. For all of the other seven (lower signal/noise) spectra, it sufficed to adopt a model consisting of a host galaxy template plus a linear function; given the weakness of the latter component, it cannot be definitely ascribed to AGN emission. The host galaxy is modeled using the S0 template from \cite{mannucci01}.

\item Narrow lines: To fully fit the [\ion{O}{iii}]$\lambda\lambda$4959,5007 doublet, depending on the S/N we required one (all spectra except the \#1 Keck and the first \#2 LDT spectra) or two (for \#1 Keck and the first \#2 LDT spectra) moderately broad (FWHM $\sim$ 1000~km s$^{-1}$) and blueshifted components in addition to the core narrow lines (Fig.~\ref{fig:keck_spec}b); these broad components are likely the product of an outflowing wind in the NLR. However, we do not adopt extra blue component for the [\ion{N}{II}] emission line doublet superposed on the broad H$\alpha$ to avoid over-parameterization, as we already obtained optimum fits considering narrow line profiles only. For the low S/N [\ion{S}{II}] line upon inspection we found some asymmetries towards the blue wavelengths in the doublets for only a few spectra (most prominent in \#1 LDT and \#2 LDT) but they are not persistent and did not exhibit a consistent trend in relation to the continuum. Furthermore, we do not observe such doublet in the best quality Keck spectrum. We thus do not adopt any additional components for [\ion{S}{II}] in any spectra. The narrow H$\beta$ emission line is extremely weak and can be detected robustly only in the Keck spectrum (\#1). We also find a narrow line feature at $\lambda \simeq 4893$~$\AA$ in the Keck spectrum (\#1) and the first two LDT spectra (\#2--3). This line feature is at a wavelength consistent with [\ion{Fe}{vii}]$\lambda$4893. However, we caution the reader that this identification is tentative, as such a line typically does not appear in isolation and is accompanied by other coronal lines such as [\ion{Fe}{vii}] $\lambda$3759, 4893, 5159, 5276, 5720, and 6086 \citep{rose2011}, which are not detected in J0428$-$00. All errors on integrated narrow line fluxes are calculated from the least squares fitting method.

\item Broad emission lines: Broad H$\beta$ emission is detected in all of the 2020--2023 spectra; a model lacking broad H$\beta$ is a worse fit as per the Akaike Information Criterion \citep{akaike1974}. In the relatively lower signal/noise spectra (\#5, 6, 8, and 10), a single broad Gaussian fit sufficed to describe the broad H$\alpha$ profile, and use of a double Gaussian provided no additional fit improvement. For all other (higher signal/noise) spectra, a double-Gaussian model was necessary to achieve the best description of the profile, and consistent with J0428$-$00's being a double peak emitter as reported by \citet{ward2024} \footnote{Given the varying levels of signal/noise across our spectra, we are not claiming that the broad H$\beta$ line intrinsically changes profile between the single or double Gaussian shapes; it is simply of matter of preferring the simpler model when it provides an adequate fit.}. Meanwhile, the broad H$\alpha$ line was fit well with a double-Gaussian profile. Parameter uncertainties for the broad components were estimated by the Maximum Likelihood method. The best fits to the H$\beta$ and H$\alpha$ profiles in the Keck spectrum (\#1) are plotted in Fig.~\ref{fig:keck_spec}.

\end{itemize}

Model fit results are summarized in Tables~\ref{tab:spec_comp1} and \ref{tab:spec_comp2}. The best fit to the Keck spectrum (\#1) is plotted in Fig.~\ref{fig:keck_spec} to illustrate the profile decompositions.
 
The resulting light-curves of broad H$\beta$ and H$\alpha$ fluxes across our campaign are plotted in the bottom panel of Fig.~\ref{fig:all_lc}. H$\beta$ and H$\alpha$ fluxes decay roughly in concert, with maximum/minimum values
of $5.8\pm0.2$ for H$\beta$ and $4.2 \pm 0.2$ for H$\alpha$. Values of fractional variability amplitude $F_{\rm var}$ are $60 \pm 6\%$ for H$\beta$ and $41 \pm 2\%$ for H$\alpha$. Both broad Balmer fluxes correlate well with X-ray and UV continuum fluxes:
zero-lag ICF values are
0.865 (X-ray--H$\beta$),
0.837 (UVW2--H$\beta$),
0.859 (UVM2--H$\beta$),
0.835 (X-ray--H$\alpha$),
0.701 (UVW2--H$\alpha$), and
0.722 (UVM2--H$\alpha$).
Again, though, data sampling precludes any meaningful search for lags or leads.

We also model the broad Balmer profiles in the archival 6dF spectrum from 2005. Broad H$\alpha$ emission is detected, and a double-Gaussian model provides a superior fit to a single-Gaussian model. However, broad H$\beta$ emission is not detected; the additions of neither single- nor double-Gaussian models yield any improvement to the fit. The spectral decomposition is displayed in Fig.~\ref{fig:6df_spec}. Our analysis support the source's classification as a type 1.9 in the 6dF spectrum \footnote{Our analysis contradicts the claim of detection of H$\beta$ by \cite{chen2022} in the 6dF spectrum. In addition, the implied Balmer decrement value obtained by \citet{chen2022} for the 6dF spectrum, with H$\alpha$ flux being roughly a third that of H$\beta$, is not physically feasible in AGN.} and supports that the source underwent an optical spectral type change from 2005 to 2020 \citep{frederick2020atel}. However, the 6dF spectrum is not flux calibrated, thus direct comparison of flux estimates with the recently-obtained spectra is not possible.

\begin{figure}
\centering
\includegraphics[scale=0.6]{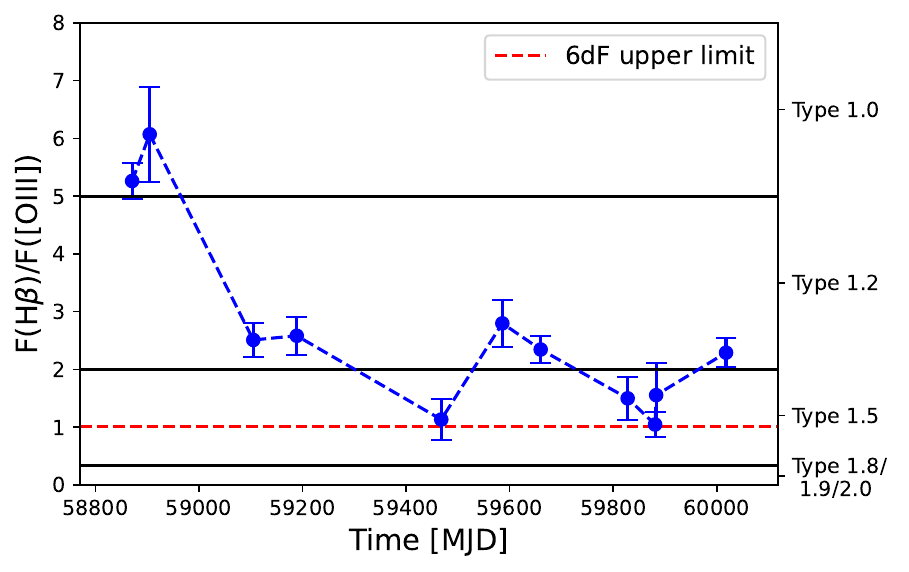}
\caption{The ratio of broad H$\beta$ flux to narrow 
[\ion{O}{iii}] $\lambda$5007 flux,
$R_{\rm H\beta/[\ion{O}{iii}]}$, as a function of time for the 2020--2023 spectra.  The upper limit to $R_{\rm H\beta/[\ion{O}{iii}]}$ from the archival 6dF spectrum is plotted as the red dashed line. The solid horizontal lines denote the boundaries between Seyfert sub-types, as per \citet{winkler1992}.} \label{fig:R_subtype}
\end{figure} 

To further quantify the evolution in the H$\beta$ profile across all spectra, including 6dF, we consider the ratio of broad H$\beta$ flux to the sum of the fluxes in the [\ion{O}{iii}] $\lambda$5007 profile, as defined in \citet{winkler1992}: $R_{\rm H\beta/[\ion{O}{iii}]} \equiv F({\rm H\beta})/F([{\rm \ion{O}{iii}}])$. In Fig.~\ref{fig:R_subtype}, we plot the resulting values of $R_{\rm H\beta/[\ion{O}{iii}]}$ as a function of time for the 2020--2023 spectra. As broad H$\beta$ emission is not detected in the 6dF spectrum, the resulting upper limit on $R_{\rm H\beta/[\ion{O}{iii}]}$ is $\sim$1.0, which is plotted as the red dashed line. We can see that $R_{\rm H\beta/[\ion{O}{iii}]}$ jumped from $<1.0$ in 2005 to values near 5--6 in early 2020, then faded to values between 1.0 and 2.8 for the remainder of the campaign.

We can also use $R_{\rm H\beta/[\ion{O}{iii}]}$ to assign approximate Seyfert subtypes to all spectra, and track J0428$-$00's subtype evolution. Following \citet{winkler1992}, sub-types 1.0, 1.2, 1.5 and 1.8 correspond to 
$R_{\rm H\beta/[\ion{O}{iii}]}$ > 5,
$2 < R_{\rm H\beta/[\ion{O}{iii}]} < 5$,
$0.33 < R_{\rm H\beta/[\ion{O}{iii}]} < 2$, and
$R_{\rm H\beta/[\ion{O}{iii}]} < 0.33$, respectively;
these regimes are marked in Fig.~\ref{fig:R_subtype}. We classify the 6dF spectrum as subtype 1.9, since broad H$\beta$ is not detected, while H$\alpha$ clearly is. The early 2020 spectra, Keck (\#1) and the first LDT spectrum (\#2) are subtype 1.0. The other nine spectra, from late 2020 through 2023, are all subtypes 1.2--1.5. However, we note two caveats regarding the use of $R_{\rm H\beta/[\ion{O}{iii}]}$ as a generic quantity for sub-type classification, which can introduce mild uncertainties into the precise subtyping. With regard to comparing different objects and/or when different instruments are used, (a) intrinsically, different objects will span different ranges of NLR characteristics and host galaxy star formation activity. (b) differences in configuration (e.g., aperture) from one instrument to the next can lead to different measurements of [\ion{O}{iii}], host galaxy light, etc. Nevertheless, for J0428$-$00, the evolution in its sub-type classification reinforces the significant evolution in the relative line strengths we have observed, consistent with its status as a changing look AGN.

\subsubsection{Testing the physical diskline model}\label{sec:diskline}

\begin{table*}
\centering
\caption{Best-fitting parameters from diskline fits to the broad Balmer profiles in the Keck spectrum}\label{tab:diskline}
\begin{tabular}{c|cccc|cc}
\hline  Line &   \multicolumn{4}{c|}{Diskline} & \multicolumn{2}{c}{Gaussian component} \\
  &    $R_{\rm in}$& $R_{\rm out}$ & Inclination ($\theta_{\rm i}$) & $\sigma_{\rm 0}$ & $c\frac{\Delta \lambda}{\lambda}$ & $c\frac{\sigma_{\lambda}}{\lambda}$ \\
& ($R_{\rm g}$)& ($R_{\rm g}$)& ($^{\circ}$) & (km s$^{-1}$)& (km s$^{-1}$) & (km s$^{-1}$)\\
\hline
H$\alpha$   & \textbf{$390 \pm 10$}  & \textbf{$1290 \pm 140$}  & \textbf{$13.2 \pm 0.2$}  & \textbf{$1160 \pm 60$} &  \textbf{$21 \pm 165$}  & \textbf{$3190 \pm 90$}\\
H$\beta$  & \textbf{$716\pm44$} & \textbf{$1000^\dag$}  & \textbf{$17.0 \pm 0.6$} & \textbf{$1000 \pm 46$} & \textbf{$290\pm150$}  & \textbf{$3600 \pm 200$}\\
\hline
\end{tabular}
\tablefoot{Best-fitting parameter values for diskline fits to the broad H$\beta$ and H$\alpha$ lines in the Keck spectrum, for the diskline + broad Gaussian model, and where fit parameters are not tied between H$\beta$ and H$\alpha$. The inner and outer radii $R_{\rm in}$ and $R_{\rm out}$ are in units of the gravitational radius ($R_{\rm g}$). The inclination angle ($\theta_{\rm i}$) is in degrees. The parameter $\sigma_{0}$ is the broadening parameter. The displacement of the peak ($c\frac{\Delta \lambda}{\lambda}$) with respect to the rest wavelength and the width of the additional Gaussian is reported in km s$^{-1}$.
The error bounds marked with $\dag$ are pegged at set lower bounds (see Sect.~\ref{sec:diskline} for details) respectively.}
\end{table*}

\begin{figure*}
\centering
\includegraphics[scale=0.44]{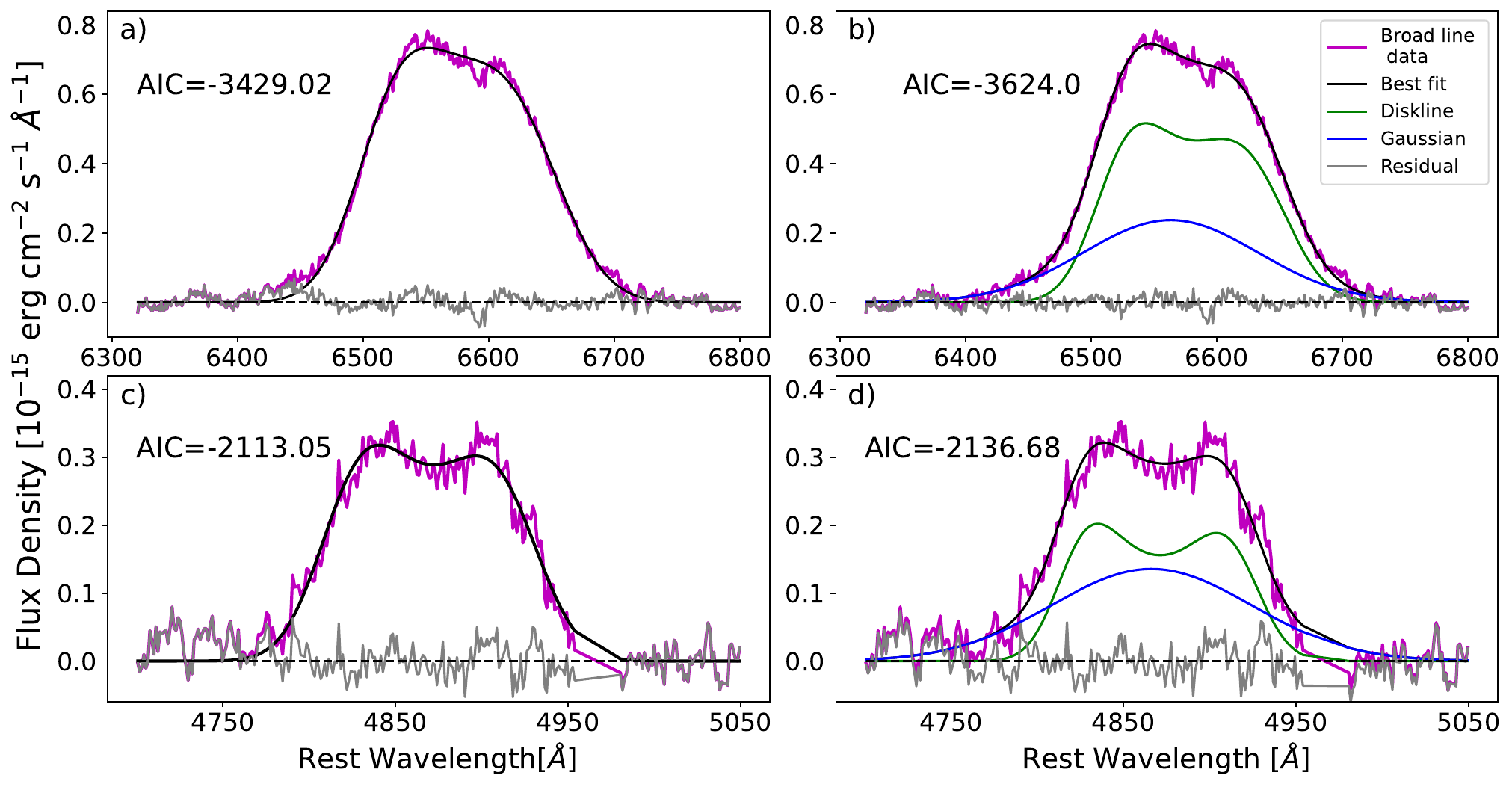}
\caption{Diskline model fits to the broad H$\alpha$ and H$\beta$ emission line profiles as described in Section \ref{sec:diskline}. 
(a) A single diskline fit to the broad H$\alpha$ emission line. 
(b) A diskline + broad Gaussian fit to the broad H$\alpha$ emission line; residuals are improved compared to panel (a). 
(c) Diskline-only best fit for H$\beta$, with parameters independent of those for the H$\alpha$ fit. 
(d) Diskline + Gaussian best fit to H$\beta$, again with parameters independent of those for the H$\alpha$; this is our best-fitting model for the broad H$\beta$ profile.
}\label{fig:diskline}
\end{figure*}

For the best-quality spectrum, that taken with Keck, we test if a physically-motivated double-peaked profile from a disk emitter \citep{chen1989a,chen1989} explains the observed H$\alpha$ and H$\beta$ profiles. This model (henceforth "diskline") assumes that the line-emitting matter is in Keplerian motion in a flattened geometry and it incorporates the effects of Doppler shifts and gravitational redshift. Following \citet{chen1989}, we apply local broadening due to electron scattering in a photoionized atmosphere \citep{halpern1984, shields1981}. The emissivity profile assumes isotropic illumination from the center.

We isolate the broad emission lines by subtracting the underlying continuum and best fit narrow profiles obtained from the phenomenological Gaussian fitting (Section \ref{subsec:2-gauss}). We tested a model with only a diskline component (A), as well as a model with a diskline + a broad Gaussian component (B). For both H$\alpha$ and H$\beta$, the Akaike Information Criterion indicates a strong preference for model B: the addition of the broad Gaussian caused the value of AIC to drop by over 15 and improved data--model residuals, as plotted in Fig.~\ref{fig:diskline}. The center of the broad Gaussian was consistent with a zero-velocity offset.

We list the best-fitting model parameters for model B in Table~\ref{tab:diskline}. The best-fitting geometrical parameters from H$\alpha$ are consistent with those obtained for multiple double peaked emitters by \citet{ward2024} using a similar diskline model. For all fits we obtained a consistent value of inclination, $\theta_{i} \simeq 15^{\circ}$. Overall, our analysis indicates that the H$\beta$ and H$\alpha$ emission lines contain similar emission components.
 
\subsection{Diagnostics of historical AGN activity}\label{sec:bpt}
\begin{figure*}
\centering
\includegraphics[scale=0.49]{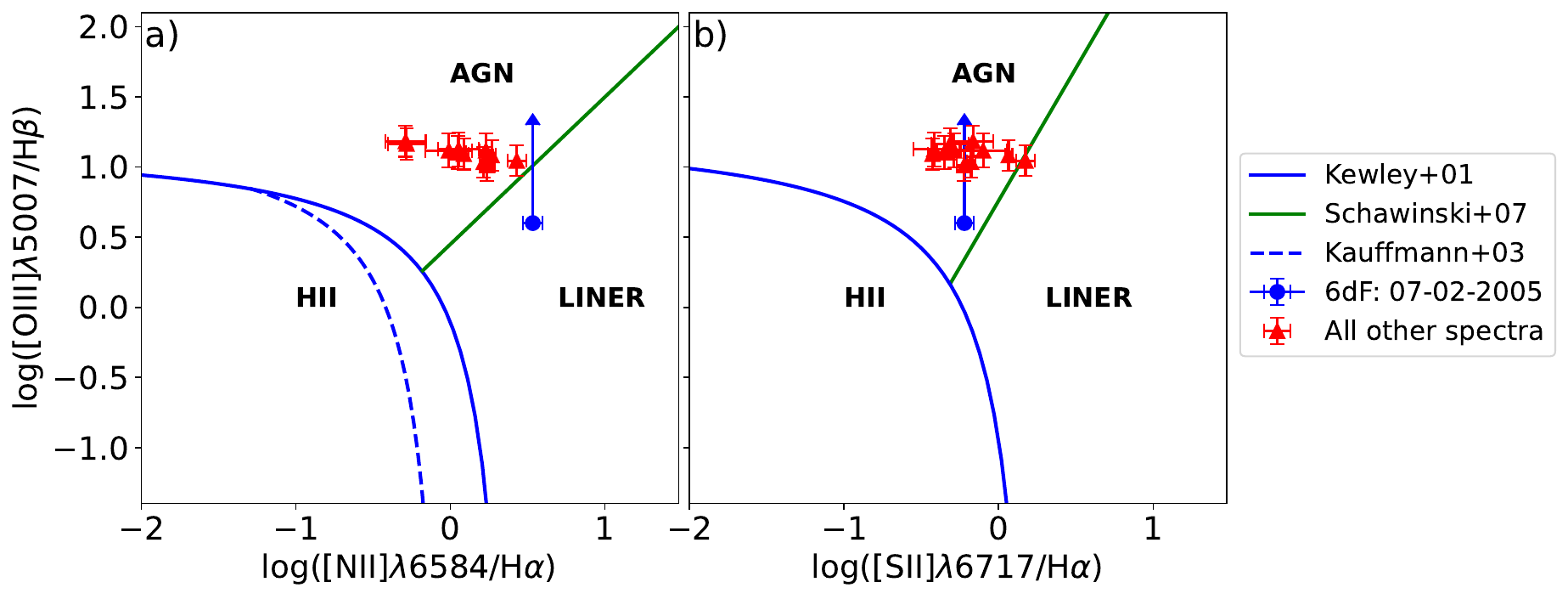}
          \caption{BPT diagrams \citep{baldwin1981} to assess activity in J0428$-$00. The line ratios (panel a) [\ion{N}{ii}]$\lambda$6583/H$\alpha$ and (panel b) [\ion{S}{ii}]$\lambda$6717/H$\alpha$ are plotted against the line ratio [\ion{O}{iii}]$\lambda$5007/H$\beta$. The classification curves are from \cite{kewley2001}, \cite{kauffmann2003}, and \cite{schawinski2007}. Most points for J0428$-$00 are clustered in the `AGN' region of the diagram, indicative of persistent AGN activity.}\label{fig:bpt}
\end{figure*}

We diagnose the historical AGN activity in J0428$-$00 via Baldwin, Phillips, and Terlevich (BPT) diagrams \citep{baldwin1981} using the best-fitting Gaussian parameters to the narrow lines.
We plot the flux ratios for both [\ion{O}{iii}]$\lambda$5007/H$\beta$ versus [\ion{N}{ii}]$\lambda$6584/H$\alpha$ and [\ion{O}{iii}]$\lambda$5007/H$\beta$ versus [\ion{S}{ii}]$\lambda$6717/H$\alpha$ along with classification curves from \citet{schawinski2007}, \citet{kewley2001}, and \citet{kauffmann2003} in Fig.~\ref{fig:bpt}.
In all spectra except the Keck spectrum, the narrow H$\beta$ line had low signal/noise and its flux was not constrained, so we adopted the narrow H$\beta$ line flux from the Keck spectrum for all spectra.
For most of the observations, the [\ion{O}{iii}]$\lambda$5007/H$\beta$ line flux ratio is high enough to position the object in the `AGN region'  of the BPT diagram (the region above the solid blue curve in Fig.~\ref{fig:bpt}), indicating
that the nucleus of J0428$-$00 has been in a prolonged ($\sim$ $10^{3}$--$10^{4}$~years) persistently accreting state.

\subsection{Estimation of black hole mass and Eddington ratio} \label{disc:blackhole_mass}
We estimate the black hole mass by using standard empirical relations between $M_{\rm BH}$, FWHM velocity, and $\lambda L_{\rm 5100}$, calibrated against nearby Seyferts whose black hole masses have been measured via BLR reverberation mapping. 
For J0428$-$00 we use the best signal/noise spectrum -- from Keck -- to estimate the above quantities. 
For the Keck spectrum, the best-fitting FWHM velocity of H$\beta$ is 8140~km~s$^{-1}$ and $\lambda L_{\rm 5100} = 4.62 \times 10^{43}$~erg~s$^{-1}$.
We use Eq.~4 from from \cite{vestergaard2006} for the H$\beta$ broad line, and obtain the an estimate $M_{\rm BH} = 3.3^{+5.6}_{-2.1} \times 10^{8} M_{\rm \odot}$. 

We also use the radius-luminosity relation between $\lambda L_{5100}$ and BLR radius ($R_{\rm BLR}$) along with the H$\beta$ FWHM to estimate black hole mass using the scaling relation from \cite{bentz09}: $M_{\rm BH} = f_{\rm vir} R_{\rm BLR} \rm{FWHM}_{\rm H\beta}^2/G$. We use the average inclination obtained from the diskline fits ($15.7^{\circ}$) to calculate the virial factor following \cite{mejia-restrepo2018}:
\begin{equation}
    f_{\rm vir} =   [4  ( \sin^2(i)  + (H/R)^2)]^{-1}
\end{equation}
For assumed values of the aspect ratio $H/R$ spanning 0.01--0.10, we obtain values of $f_{\rm vir}$ spanning 3.2--3.7 respectively.
However, \cite{mejia-restrepo2018} and \cite{yu2019} also indicate that the virial factor is inversely proportional to the FWHM of the line and could be as low as 0.5 for FWHM values as high as $8000$~km~s$^{-1}$.
Thus for values of $f_{\rm vir}$=0.5 or 3.7 we obtain black hole masses of $M_{\rm BH} = 1.3^{+0.5}_{-0.3}\times10^8 M_{\rm \odot}$ and $8.7^{+3.0}_{-2.3}\times10^8 M_{\rm \odot}$, respectively. We adopt the average of all these estimates, $4\times10^8 M_{\rm \odot}$, for all calculations. However, as a caveat, the standard relations used below are based on persistently-accreting Seyferts, while J0428$-$00 was monitored while undergoing a sudden transient luminosity flare.

\section{Discussion} \label{sec:disc}
\subsection{Summary of main observations}
The source exhibited a sharp flare in the optical continuum in early 2020, with a flux increase over 35 days, followed by a decay that we could track for an additional $\sim$30 days before the sun gap.
Meanwhile, the optical spectrum underwent a transition: in the 2005 spectrum, broad H$\beta$ emission was not detected (sub-type 1.9).
In the spectra taken during Jan.--Feb.\ 2020, broad H$\beta$ emission was present and strong (Fig.~\ref{fig:all_spec}), consistent with subtype 1.0 (Sec \ref{subsec:2-gauss}).
The profiles of broad H$\alpha$ and H$\beta$ are fit well by a double-peaked profile.
After the flare peak in the optical, we tracked the subsequent decay in the optical, UV, and X-ray continua until 2023. The broad H$\alpha$ and H$\beta$ emission line fluxes track the long-term continuum (Fig.~\ref{fig:all_lc}).

The X-ray spectra are fit well by a single unabsorbed power law with a flat photon index ($\sim$1.9).
The X-ray and UV continua decay in concert, and the broadband SED (Fig.~\ref{fig:broad_band}) is consistent with the X-rays being generated by thermal Comptonization of low-energy seed photons from the accretion disk.
Intriguingly, the soft X-ray excess, a near-ubiquitous feature of the X-ray spectra of nearby Seyferts, is absent in J0428$-$00.

J0428$-$00 exhibits persistent narrow emission lines of [\ion{O}{III}]$\lambda\lambda$4959,5007, [\ion{N}{II}]$\lambda\lambda$6548,6584, and [\ion{S}{II}]$\lambda\lambda$6716,6731. The resulting BPT-diagnostic ratios (Fig.~\ref{fig:bpt}) indicate ongoing AGN activity for of order millennia.

Finally, the \textit{WISE} IR lightcurves exhibit a broad flare (Fig.~\ref{fig:all_lc}), with the peak in W2 delayed by roughly 170 days with respect to the main optical flare, consistent with a "dust echo" in circumnuclear warm dust. 

Below, we construct the "big picture" of how the temporary continuum flaring is connected to the optical spectral changing-look transition, the X-ray emission, and the IR flux peak, while we speculate on the underlying mechanism(s) responsible for the flare.

\subsection{Characteristics of the transient}
The nature of a supermassive black hole transient flaring event can be consistent with either AGN-like accretion, or with a tidal disruption event (TDE) of a star by a quiescent black hole \citep{rees1988,evans1989}. 
While TDEs occur can occur in quiescent galaxies, recent observations have uncovered some nuclear transients that display attributes consistent with both TDE- and AGN-like accretion, suggesting that both channels of accretion can occur simultaneously \citep[e.g.,][]{ricci2020,holoien2022,homan2023}. 

The X-ray spectra of J0428$-$00 are Seyfert-like with a photon index $\Gamma_{\rm X}$ remaining near 1.9, and we do not observe any drastic variation in photon index. This hard X-ray spectral behavior \citep{auchettl2018} is inconsistent with that demonstrated by thermal TDEs, which typically exhibit much softer X-ray spectra \cite[$\Gamma_{\rm X}$ $>$ 3, typically; e.g.,][]{gezari2021}. As a caveat, though, the hard X-ray photon index distribution of TDEs is broad, and can sometimes encompass such low values \citep[e.g., with $\sim0.3$ probability as per][]{auchettl2018}.

Other observations in J0428$-$00 instead support AGN-like accretion: A persistent broad H$\alpha$ line was seen in both the archival 6dF spectra as well as during the recent campaign, indicative of AGN activity both in 2005 and during 2020--2023. This is supported also by the narrow-line BPT diagnostics (Figure \ref{fig:bpt}), which classify this source as an AGN with prolonged accretion activity.

The ratio of the luminosity of the hard X-ray luminosity to the luminosity of the [\ion{O}{III}]$\lambda$5007 line can also discriminate between persistent AGN activity over the last $\sim$ $10^3$--$10^4$ years versus a TDE event \citep{heckman2005}. \citet{sazonov2021} found this ratio to be around $10^2$ in a sample of local AGNs and around $10^4$ in a sample of TDEs. For J0428$-$00, we estimate the ratio $R_{\rm X/[\ion{O}{III}]}$ of $L_{\rm 0.2-6.0~keV}/L_{[\ion{O}{III}]}$ following \cite{sazonov2021}. At flare peak, when the X-ray luminosity highest, $R_{\rm X/[\ion{O}{III}]}$ is $10^{2.8}$. Three years after the peak has passed $R_{\rm X/[\ion{O}{III}]}$ is $10^{2.0}$. The high value of $R_{\rm X/[\ion{O}{III}]}$ near the flare peak can be interpreted as being due to a rapid but temporary increase in X-ray flux during the outburst while the [\ion{O}{III}]$\lambda$5007 line flux remained static. 

To summarize, the observed activity in J0428$-$00 seems more reminiscent of a temporary change in accretion rate in a disk-like accretion flow in a persistently-accreting AGN than accretion from a tidally-disrupted star.
Below we consider the emission and variability properties in the context of AGN accretion.

\subsection{Origin of the X-ray continuum}\label{disc:xray}
As noted above the X-ray photon index $\Gamma_{\rm X}$ stayed near roughly $1.9$ as the multiband continuum flaring decayed (Section \ref{sec:xray-analysis} and Table~\ref{tab:x-ray}). This value is consistent with hot-Comptonization of seed photons from an accretion disk is a hot corona \citep{haardt1991,haardt1993} in Seyfert galaxies \citep[e.g.,][]{mateos2010}. We find no evidence for significant X-ray spectral variability over three years despite strong X-ray continuum flux variability. In principle, $\Gamma_{\rm X}$ is dependent on the ratio of heating \citep[e.g., from magnetic processes;][]{balbus1998} to cooling (from up-scattering of seed photons) in the corona \citep[Eq.14 in][]{beloborodov1999}. It is conceivable that in J0428$-$00, (a) the total energy injected into the corona by the incident seed remained less than the total internal energy of the corona during the continuum decay, and/or (b) coronal parameters such as optical depth or geometry remained stable over time, and flux variations in the X-rays resulted from seed photon variations, albeit with no coronal cooling.

CLAGNs display a wide range of X-ray variability properties across multiple sources. For example, NGC 1566 exhibited an outburst wherein spectral evolution resembles a q-diagram in the hardness-intensity plane \citep{jana2021}. Its X-ray photon index is qualitatively consistent with the $\Gamma_{\rm X}$--flux relations observed in normal Seyferts and Black Hole X-ray Binaries (BHXRBs) \citep[e.g.][]{papadakis2009} and hardness intensity characteristics of BHXRBs \citep[e.g.][]{remillard2006}. In contrast, two X-ray spectra taken two years apart in the CLAGN IRAS 23226$-$3843 \citep{kollatschny2023} exhibit variability only in X-ray normalization following an outburst, with no significant evolution in X-ray spectral shape. Such behavior is qualitatively similar to that seen following the flare in J0428$-$00. 

A soft X-ray excess, manifested as a very steep power-law-like emission below roughly 1--2~keV, is near-ubiquitous in normal Seyferts \citep{halpern1984,turner1989}. The absence of a soft excess in the X-ray spectra of J0428$-$00 is puzzling. Its origin in normal Seyferts is debated \citep[e.g.,][]{sobolewska2007}. The leading models are thermal Comptonization from a warm ($k_{\rm B}T_{\rm e} \sim 1$~keV), optically thick ($\tau \gtrsim 10$) corona \citep{mehdipour2011,done2012, petrucci2018} or blurred reflected emission from the innermost, ionized accretion disk \citep{crummy2006, garcia2013}. In the first case, the lack of an observed soft excess above 0.2 keV could mean that the warm corona is simply physically absent in J0428$-$00. Alternately, our broadband SED fits (Section \ref{sec:broadband_analysis}) using AGNSED suggest that, if there is a warm corona present, its emission remains localized to energies below $0.2$~keV, while all emission above 0.2 keV is Comptonization from the a hot corona. Given the range of accretion rates for J0428$-$00 inferred during our campaign, log($\lambda_{\rm Edd}$) $\sim$ $-$2 to $-$3, the lack of soft excess is consistent with the findings of \citet{hagen24}, that the soft excess is typically not present in the X-ray spectra of sources accreting at values of log($\lambda_{\rm Edd}$) below roughly $-$1.5. Meanwhile, in the case of blurred reflection, such a component might be absent because of the absence of an optically-thick, cold, accretion disk in the innermost regions of the accretion flow, leading to inefficient illumination by, and reflection of, the X-ray continuum photons. A final possibility is that such a disk component exists but is over-ionized.

\begin{figure}[ht]
\centering
\includegraphics[scale=0.65]{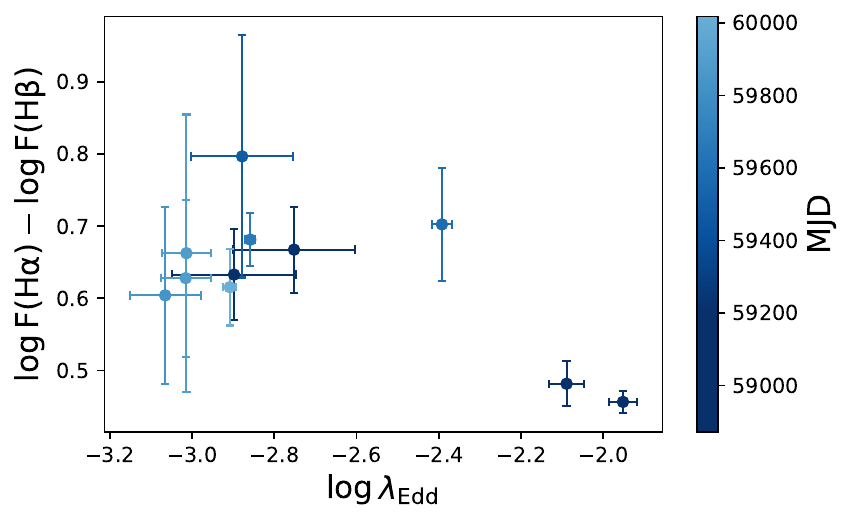} 
\caption{Balmer decrement as a function of accretion rate relative to Eddington for our 2020--2023 optical spectra; the increase in Balmer decrement as luminosity fades from 2020 to 2023 is qualitatively consistent with the expected response to the decrease in ionizing luminosity \citep{wu2023}. The shades on the markers indicate time: deeper shade indicates earlier in time.}\label{fig:BalmDecr}
\end{figure}

\subsection{The stability of the accretion structure} \label{disc:comparison}
Major changes in the overall disk/corona structure in AGN/quasars can be tracked via the UV-to-X-ray spectra index $\alpha_{\rm OX}$ \citep{tananbaum1979}. Evolution in $\alpha_{\rm OX}$ as a function of $\lambda_{\rm Edd}$ has been recorded both across samples of normally varying quasars \citep[e.g.,][]{lusso10} and in individual changing-look quasars \citep{ruan19}. For J0428$-$00, we obtain those values from the SED fits in Sect.~\ref{sec:broadband_analysis}. The resulting values of $\alpha_{\rm OX}$ during our campaign span from from 1.1 to 1.4, averaging near 1.2. The scatter in $\alpha_{\rm OX}$ is due to the short-term variability in both the UV and especially the X-ray bands, but $\alpha_{\rm OX}$ does not exhibit anu systematic trends precluding any major changes in global disk/corona structure. These values of $\alpha_{\rm OX}$ are similar to those for X-ray-selected type-1 quasars at the same UV luminosity, as reported in \cite{lusso10}.

\subsection{Broad Balmer lines and emitter geometry} \label{disc:broadline}
The broad H$\alpha$ emission line is persistent in all stages of observation, including in 2005, whereas the broad H$\beta$ only appeared during the 2020--2023 transient event, displaying much stronger variability than the H$\alpha$ line (Fig.~\ref{fig:all_lc}f).\cite{wu2023} demonstrated that major changes in the Balmer decrement can be driven by changes in the incident ionization continuum due to a change in accretion rate. Such a change in ionizing flux affects the responsivity of the H$\beta$ emission line more than the H$\alpha$ line due to the $Q$-dependent opacity in the emitter \cite[details are given in Section 5 of][]{wu2023}. This can potentially lead to large relative changes in the broad H$\beta$ line strength in intermediate-type Seyferts, potentially driving a CLAGN transition. In Fig.~\ref{fig:BalmDecr}, we plot the Balmer decrement as a function of $\lambda_{\rm Edd}$ for our 2020--2023 optical spectra. For each optical spectrum, we estimate $\lambda_{\rm Edd}$ based on the X-ray flux from the X-ray observations nearest in time, and using bolometric corrections from \citet{Duras2020}. As the continuum flaring fades, the Balmer decrement increases substantially (factor of 2.2); this behavior is qualitatively consistent with the theoretical predictions of \citet{wu2023}, as well as with the empirical behavior of the Seyferts/CLAGN discussed therein.

Furthermore, the optical broad lines are double peaked, where the high signal-to-noise Keck spectrum can explained by a diskline component that extends from around 500~$R_{\rm g}$ to 1000 $R_{\rm g}$, consistent with \cite{ward2024}. It is a general trend that double peaked Balmer line profiles are found prevalently in low luminosity AGNs \citep{elitzur2009,elitzur2014}. Similar trends are predicted by the dynamic Failed Radiatively Accelerated Dusty Outflow \citep[FRADO,][]{naddaf2021,naddaf2022} for values of Eddington ratio ($\lambda_{\rm Edd}$) below roughly $10^{-2}$). J0428$-$00 broadly exhibits an Eddington ratio $< 10^{-2}$ near its peak, and the observed profile of its broad emission lines is thus consistent with the above-mentioned trends.

\subsection{A possible origin for the flare}
Constraints on the mechanism driving the luminosity flare can be derived from comparing the observed flaring timescale ($\gtrsim$ 1 month) to standard timescales of variability associated with a accreting disk of matter \cite[e.g.][]{treves1988}. The standard timescales for AGN variability are light-crossing ($t_{\rm lc} = R/c$), dynamic ($t_{\rm dyn} = \sqrt{R^3/GM_{\rm BH}}$), thermal ($t_{\rm th} = (1/\alpha)t_{\rm dyn}$), and viscous ($t_{\rm visc} =(H/R)^{-2}t_{\rm th}$) timescale, where $\alpha$ and $H$ are viscosity parameter and disk thickness respectively. At arbitrary radii of $R \sim$ 10--100 $R_{\rm g}$, assuming a viscosity parameter $\alpha$ of 0.1, a scale height $H/R$ of $10^{-3}$ (for a thin disk) and 1 (for a thick disk), and a black hole mass of $M_{\rm BH}=4\times10^8 M_{\rm \odot}$, we find that the observed flare (roughly 35-day increase) is most compatible with the dyanmical and thermal timescales, and the viscous timescale for a geometrically-thick disk.

We posit that the flare in J0428$-$00 is compatible with, among other models, radiation pressure-driven disk instability models as described in \citet{lightman_eardley1974} and e.g., \citet{sniegowska2020}. In this model, an unstable annular region of the disk exists between a radiatively-inefficient inner region and the geometrically-thin outer disk. Initially, it is filled with tenuous gas, and new gas accretes inward from the outer disk on the viscous timescale for a thin disk; luminosity increases only gradually. Initially, gas pressure is greater than radiation pressure. However, when radiation pressure exceeds gas pressure locally, it triggers a heat wave that propagates inward on the local thermal timescale. It causes $\frac{H}{R}$ to increase, and the shortened viscous timescale drives the local accretion rate to drastically increase, leading to a burst of thermally-emitted radiation. The unstable zone accretes its gas inwards faster than gas is replenished. As cooling becomes greater than heating, the local accretion rate and luminosity subsequently fade; the cooling is dominated by the thermal timescale. Given that the rise and decay times of the flare in the optical band (roughly a month) are each broadly consistent with the thermal timescale, then a radiation pressure-driven instability can be a plausible explanation for the observed flaring in J0428$-$00.

Finally, we note that the light curves calculated from the simulations of \citet{sniegowska2020,sniegowska2023} qualitatively resemble the observed light curves in J0428$-$00 in this manner.

\section{Conclusions}
\label{sec:summary}
We report the detection of a new optical changing-look Seyfert: J0428$-$00 was originally a type-1.9 Seyfert, but in 2020 it exhibited a multi-wavelength flare captured in the optical by ZTF and in the X-ray by eROSITA's eRASS surveys. Tied to the extreme variability was a change in optical spectral classification from type 1.9 to 1, as broad H$\beta$ emission had appeared by 2020. We initiated a three-year multi-wavelength follow-up campaign to track the source's emission as the flare gradually subsided, and to determine the nature of this transient event.

The recent X-ray spectra, combined with indications of historical AGN activity, argue against the event being due to the tidal disruption of a star by a supermassive black hole. Instead, the flare is more likely the result of a disk instability in a previously-existing accretion structure. Additional key findings include:

\begin{itemize}\setlength\itemsep{0.5ex} \item J0428$-$00 exhibits narrow emission lines of
[\ion{O}{iii}]$\lambda\lambda$4959,5007,
[\ion{N}{ii}]$\lambda\lambda$6548,6584, and
[\ion{S}{ii}]$\lambda\lambda$6716,6731. The BPT ratios indicate ongoing AGN activity for of order millennia (Section \ref{sec:bpt}).

\item J0428$-$00 exhibits double-peaked broad Balmer emission lines. H$\alpha$ is relatively strong and persistent, even in the archival 6dF spectrum. H$\beta$ appears only in 2020, with $R_{\rm H\beta/[\ion{O}{iii}]}$ jumping from $<$1.0 to 5--6, consistent with being driven by the continuum flare and the sudden increase in the flux of ionizing photons. The broad Balmer line fluxes overall track the continuum, and slowly fade as the continuum flux subsides. The evolution in Balmer decrement (Fig.~\ref{fig:BalmDecr}) is qualitatively consistent with the expected response to the decay in the ionizing continuum \citep{wu2023}. The broad Balmer profiles' double peaks are fit well by models based on a disk-type geometry, extending from roughly a few hundred $R_{\rm g}$ to roughly $1000~R_{\rm g}$.

\item The X-ray spectra are fit well by a power law with a flat photon index ($\sim$ 1.9), typical for radio-quiet Seyferts. In addition, the X-ray and UV continuua decay in concert (Fig.~\ref{fig:all_lc}a and b). The X-rays are thus consistent with thermal Comptonization of low-energy seed photons. However, the relative stability in photon
index despite the strong X-ray variability could indicate relative stability in the hot corona's properties ($\tau$; geometry).

\item Intriguingly, the soft X-ray excess, a near-ubiquitous feature of the X-ray spectra of nearby Seyferts, is absent in J0428$-$00. This could indicate the absence or extreme weakness of a warm corona, in the context of warm Comptonization models (Section \ref{sec:broadband_analysis}). The lack of soft excess is consistent with the notion that the soft excess is typically missing in sources accreting at values of log($\lambda_{\rm Edd}$) below roughly $-$1.5 \citep{hagen24}.

\item The \textit{WISE} lightcurves exhibit a broad flare, with the peak in the W2 band delayed by roughly 170 days with respect to the main optical flare, consistent with reprocessing in a $\sim$pc-scale circumnuclear dust structure.
\end{itemize}

All these observations are consistent with a scenario where the main trigger of the events is a temporary instability in the accretion flow structure. A Lightman-Eardley instability triggered in the inner-disk ($\sim$50 to $100 R_{\rm g}$) is consistent with the observed timescale of the main flare from 2020. Additionally, the continuum variability light curves of J0428$-$00 qualitatively resemble those theoretically calculated in \cite{sniegowska2020} and \cite{sniegowska2023}. To summarize, this disk instability resulted in a temporary increase in the local accretion rate in the disk, yielding optical/UV thermal continuum flaring. In turn, the resulting increase in optical/UV seed photons drove the observed X-ray flaring via thermal Comptonization, and the increase in ionizing far-UV luminosity drove the increase in broad H$\beta$ flux as the pre-existing disk-like BLR was illuminated. Later, the optical/UV/X-ray flaring emission reached the dusty torus, resulting in the observed IR flare.

Time-domain astronomy is yielding an ever-increasing accumulation of peculiar supermassive black hole transient events. The discovery of flaring activity in J0428$-$00 is part of this multi-wavelength time domain effort, and includes eROSITA's new window into X-ray time domain studies. CLAGN and flaring AGN are events that yield insight into major changes in accretion rate, though sometimes those changes are temporary. Consequently, it is critical to study such events and understand the mechanism(s) driving them to fully understand AGNs' accumulated accretion histories.

\begin{acknowledgement}
The authors thank the anonymous referee for the comments and suggestions. The authors acknowledge insightful discussions with Prof.\ Chris Done, Prof.\ Ji\v{r}\'{i} Svoboda, and Prof.\ Piotr \.Zycki. TS acknowledges full and partial support from Polish Narodowym Centrum Nauki grants 2018/31/G/ST9/03224 and 2016/23/B/ST9/03123, and partial support from Deutsches Zentrum für Luft- und Raumfahrt (DLR) grant FKZ 50 OR 2004. AM acknowledges full or partial support from Polish Narodowym Centrum Nauki grants 2016/23/B/ST9/03123, 2018/31/G/ST9/03224, and 2019/35/B/ST9/03944. DH acknowledges support from DLR grant FKZ 50 OR 2003. MK is supported DLR grant FKZ 50 OR 2307. MG is supported by the EU Horizon 2020 research and innovation programme under grant agreement No 101004719. SH is supported by the German Science Foundation (DFG grant number WI 1860/14-1). DAHB \& JB acknowledge support from the National Research Foundation.

This work is based on data from eROSITA, the soft X-ray instrument aboard SRG, a joint Russian-German science mission supported by the Russian Space Agency (Roskosmos), in the interests of the Russian Academy of Sciences represented by its Space Research Institute (IKI), and the Deutsches Zentrum für Luft- und Raumfahrt (DLR). The SRG spacecraft was built by Lavochkin Association (NPOL) and its subcontractors, and is operated by NPOL with support from the Max Planck Institute for Extraterrestrial Physics (MPE). The development and construction of the eROSITA X-ray instrument was led by MPE, with contributions from the Dr. Karl Remeis Observatory Bamberg \& ECAP (FAU Erlangen-Nuernberg), the University of Hamburg Observatory, the Leibniz Institute for Astrophysics Potsdam (AIP), and the Institute for Astronomy and Astrophysics of the University of Tübingen, with the support of DLR and the Max Planck Society. The Argelander Institute for Astronomy of the University of Bonn and the Ludwig Maximilians Universität Munich also participated in the science preparation for eROSITA. The eROSITA data shown here were processed using the eSASS software system developed by the German eROSITA consortium.

This work is based on observations obtained with XMM-Newton, an ESA science mission with instruments and contributions directly funded by ESA Member States and NASA. This research has made use of data and/or software provided by the High Energy Astrophysics Science Archive Research Center (HEASARC), which is a service of the Astrophysics Science Division at NASA/GSFC.

This work has made use of data from the Asteroid Terrestrial-impact Last Alert System (ATLAS) project. The Asteroid Terrestrial-impact Last Alert System (ATLAS) project is primarily funded to search for near earth asteroids through NASA grants NN12AR55G, 80NSSC18K0284, and 80NSSC18K1575; byproducts of the NEO search include images and catalogs from the survey area. This work was partially funded by Kepler/K2 grant J1944/80NSSC19K0112 and HST GO-15889, and STFC grants ST/T000198/1 and ST/S006109/1. The ATLAS science products have been made possible through the contributions of the University of Hawaii Institute for Astronomy, the Queen’s University Belfast, the Space Telescope Science Institute, the South African Astronomical Observatory, and The Millennium Institute of Astrophysics (MAS), Chile.

Based on observations obtained with the Samuel Oschin Telescope 48-inch and the 60-inch Telescope at the Palomar Observatory as part of the Zwicky Transient Facility project. ZTF is supported by the National Science Foundation under Grant No. AST-1440341 and AST-2034437 and a collaboration including Caltech, IPAC, 
the Weizmann Institute for Science, 
the Oskar Klein Center at Stockholm University, 
the University of Maryland, 
the University of Washington,
Deutsches Elektronen-Synchrotron and Humboldt University, 
Los Alamos National Laboratories,
the TANGO Consortium of Taiwan, 
the University of Wisconsin at Milwaukee, 
Trinity College Dublin, 
Lawrence Berkeley National Laboratories,
Lawrence Livermore National Laboratories,
and IN2P3, France. 
Operations are conducted by COO, IPAC, and UW.

This paper uses observations made from the South African Astronomical Observatory (SAAO).

Some of the observations reported in this paper were obtained with the Southern African Large Telescope (SALT) under programs 2020-2-MLT-008 (PI: A.\ Markowitz) and 2021-2-LSP-001 (PI: D.\ Buckley). Polish participation in SALT is funded by grant No.\ MEiN nr 2021/WK/01. The paper is based on observations collected at the European Southern Observatory under ESO programme 109.23MH.001.

This publication makes use of data products from the Wide-field Infrared Survey Explorer, which is a joint project of the University of California, Los Angeles, and the Jet Propulsion
Laboratory/California Institute of Technology, funded by the National Aeronautics and Space Administration. This publication also makes use of data products from NEOWISE, which is a project of the Jet Propulsion Laboratory/California Institute of Technology, funded by
the Planetary Science Division of the National Aeronautics and Space Administration.

This research has made use of the NASA/IPAC Extragalactic Database (NED), which is funded by the National Aeronautics and Space Administration and operated by the California Institute of Technology.

We made extensive use of the following open-source \texttt{python} packages: \texttt{NumPy} \citep{numpy}, \texttt{Matplotlib} \citep{matplotlib}, \texttt{SciPy} \citep{scipy}, and \texttt{Astropy}\citep{astropy2022}.

\end{acknowledgement}

\bibliographystyle{aa}
\bibliography{aaJ0428-00}

\begin{appendix}

\onecolumn
\section{Scaling the spectra using the [\ion{O}{III}]$\lambda$5007 line}\label{sec:oiii-scaling}

We scale all the optical spectra using the [\ion{O}{III}]$\lambda$5007 narrow emission line using the profile fitting method similar to \cite{fausnaugh17}, and we summarize our procedure here. We  fit the [\ion{O}{III}]$\lambda$5007 line profile of each spectrum to a selected reference spectrum -- the Keck spectrum. We smooth the reference spectrum with a Gaussian kernel of width 3.5 \AA\ and then select a limited window ($5325 \AA < \lambda < 5390 \AA$, observed frame) around the [\ion{O}{III}]$\lambda$5007 line for fitting. Subsequently, we select the [\ion{O}{iii}]$\lambda$5007 line region and the blueward ($\lesssim 5340 \AA$) and the redward ($\gtrsim 5374 \AA$) continuum region for all spectra. We then subtract the continuum from the line region to isolate only the [\ion{O}{iii}]$\lambda$5007 line profile. We use GW-MCMC \cite[\texttt{emcee,}][]{goodman2010} to fit the emission line of the target spectrum, after smoothing, to the smoothed reference spectrum\footnote{\url{https://github.com/tathagatas1996/OIII_Flux_scaling_AGN}}. This methodology returns the following parameters: (a) the standard deviation ($\sigma$) of the Gaussian convolution function, (b) the flux correction/scaling factor, (c) the wavelength shift correction, which aligns the line-center of the target spectrum to match the reference spectrum. Best-fitting values of $\sigma$ were typical 0.5--2~\AA, smaller than the typical measured narrow-line widths. This process is performed on the redshifted spectra before applying extinction correction. After applying the scaling factors, the differences between the flux of the target spectra and the reference spectrum were less than 25\%. Furthermore, discrepancies can originate due to the degradation in the data quality during later phases of the monitoring, or other systematic issues as discussed in the caveat below.

As a caveat, we specify that the scaling factors can potentially be affected by relative differences in the flux calibrations between the red and blue sides. This is evident in our dataset, where we see a significantly boosted [\ion{S}{ii}]$\lambda\lambda$6716,6732 doublet in the first LDT (\#2) and third LDT (\#4) spectra. We believe that this apparent variability in [\ion{S}{ii}]$\lambda\lambda$6716,6732 flux is not intrinsic to the source, and might originate due to an overestimation of the ([\ion{O}{iii}]$\lambda$5007-based) flux scaling factor due to underestimated blueward flux in the LDT spectra due to calibration issues. However, given the closeness in wavelength of [\ion{O}{iii}]$\lambda$5007 and H$\beta$ our conclusions regarding H$\beta$ flux variability are likely secure, i.e.\ H$\beta$ still clearly tracks the continuum flux. We do not expect the estimated broad line and continuum correlation coefficients and fractional variability amplitudes (Section \ref{subsec:2-gauss})  to be affected by these issues to a significant extent such that it affects our conclusions.

\newpage
\section{Modeling the optical spectrum}\label{sec:continuum-fitting}
All the optical spectra are taken by different instruments and at different epochs hence the data have different data quality. Thus, our fitting models for local line regions differ slightly for each spectrum. We summarize the spectral components in Table~\ref{tab:spec_comp1}. In the table, n-Gaussian represents n-number of Gaussian profiles used to model a given line. In all cases except for the Keck spectrum, we could not confirm the presence of the narrow H$\beta$ line. [\ion{O}{iii}]$\lambda$5007 and broad H$\beta$ and H$\alpha$ fluxes are reported in Table~\ref{tab:spec_comp2}.

\begin{figure*}
\centering
\includegraphics[scale=0.43]{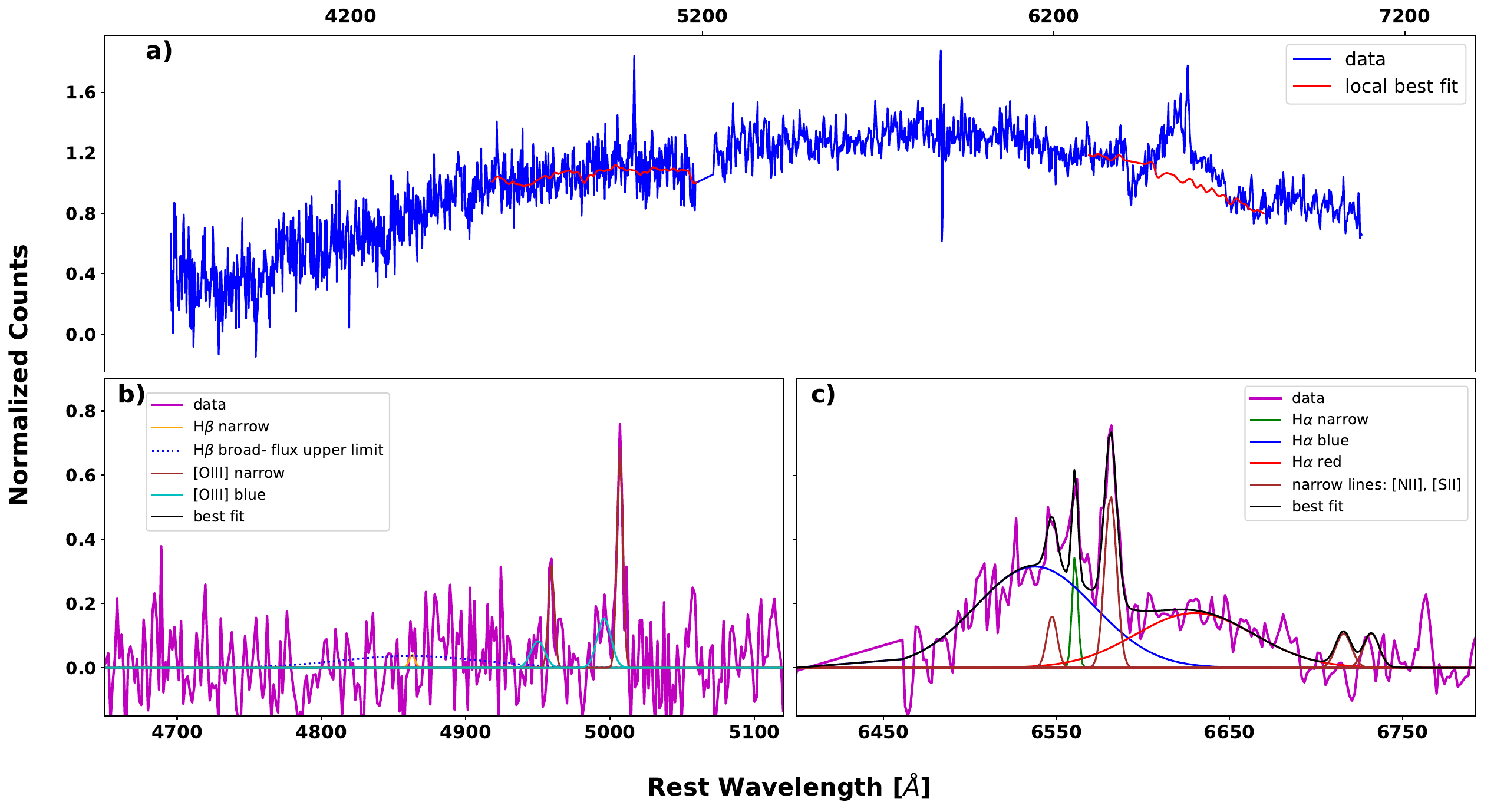} 
\caption{The mean-normalized 6dF spectrum, taken on 7th February 2005; it is not flux calibrated. A broad H$\beta$ line is not detected; the dashed black line represents the profile corresponding to the upper limit on the amplitude of a broad Gaussian centered near H$\beta$. This upper limit is comparable to the noise seen in the spectrum, thus barring us from claiming any detection of broad H$\beta$ emission.}\label{fig:6df_spec}
\end{figure*}

\begin{figure*}
\centering
\includegraphics[scale=0.43]{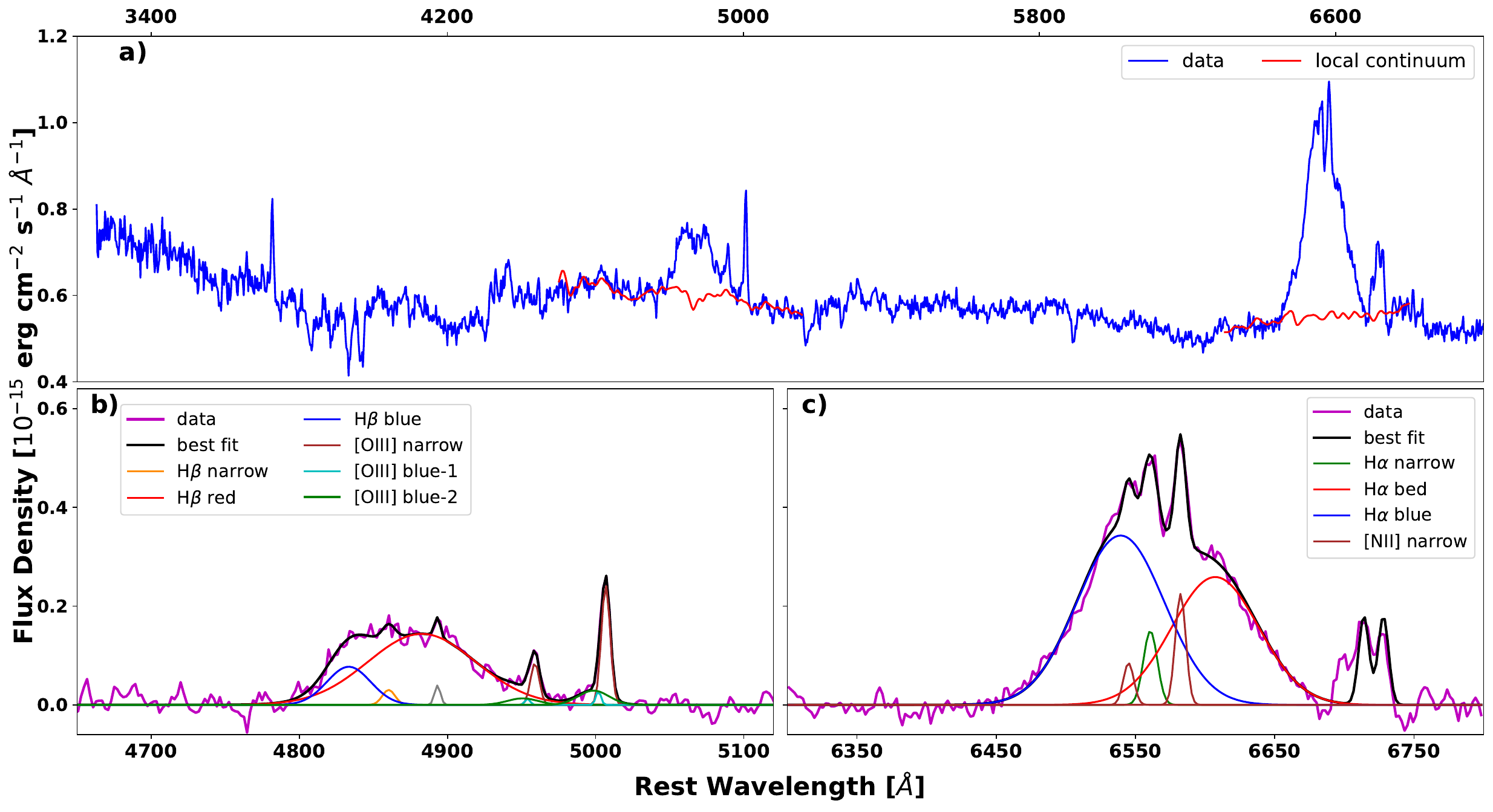} 
\caption{Same as Fig.~\ref{fig:keck_spec} but for spectrum \#2, taken at LDT on 26th February 2020.}\label{fig:dct1_spec}
\end{figure*}

\begin{figure*}
\centering
\includegraphics[scale=0.43]{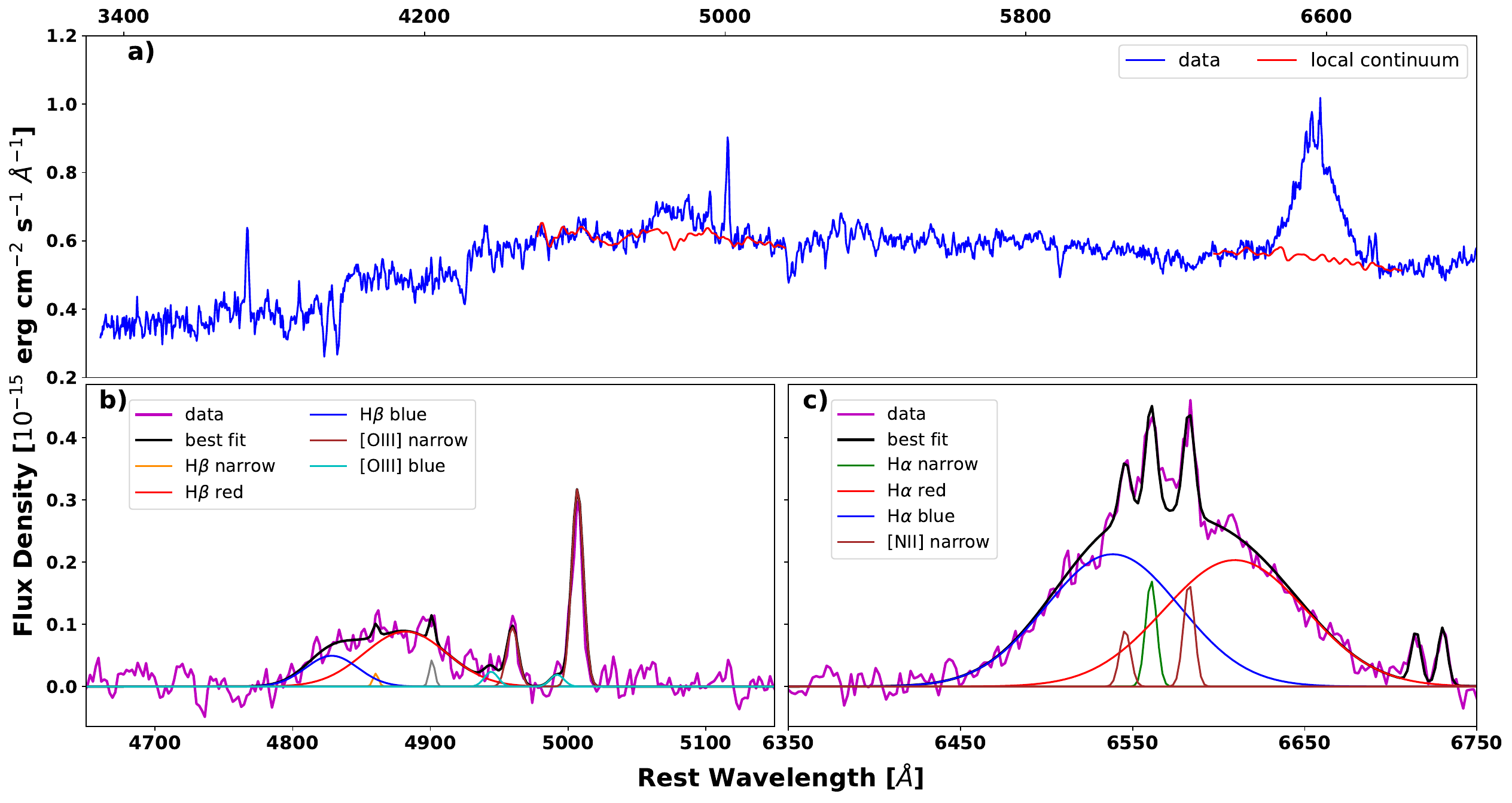} 
\caption{Same as Fig.~\ref{fig:keck_spec} but for spectrum \#3, taken at LDT on 13th September 2020.}\label{fig:dct2_spec}
\end{figure*}

\begin{figure*}
\centering
\includegraphics[scale=0.43]{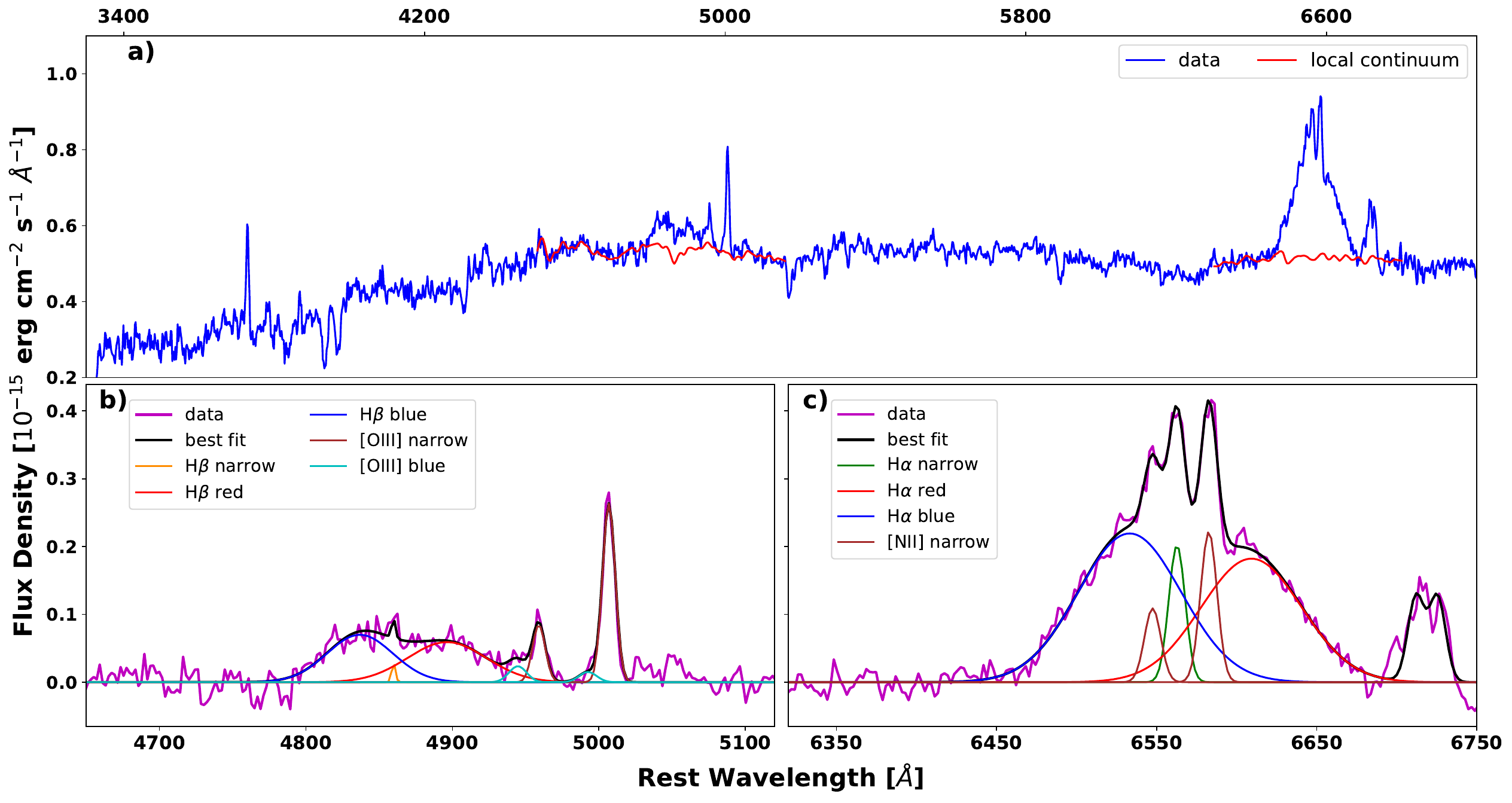} 
\caption{Same as Fig.~\ref{fig:keck_spec} but for spectrum \#4, taken at LDT on 6th December 2020.}\label{fig:dct3_spec}
\end{figure*}

The 6dF count spectrum, normalized by mean count value, is shown in Fig.~\ref{fig:6df_spec}. For brevity, we plot the best-fitting models only for selected optical spectra (\#2, \#3, and \#4) in Figs.~\ref{fig:dct1_spec}, \ref{fig:dct2_spec}, and \ref{fig:dct3_spec}.

\begin{table*}[h!]
\centering
\caption{Overview of model fits to the H$\beta$ and H$\alpha$ profiles in optical spectra.}\label{tab:spec_comp1}
\begin{tabular}{lccccccccc}
\hline
Date      & Telescope & H$\beta$-$\lambda$ range   & H$\alpha$-$\lambda$ range & [\ion{O}{III}]-model & Broad H$\beta$-model \\
\hline
53408     & 6dF       & (4600~$\AA$, 5160~$\AA$)   & (6416~$\AA$, 6760~$\AA$)     & 1-Gaussian & 2-Gaussian \\
\#1 58871 & Keck      & (4500~$\AA$, 5160~$\AA$)   & (6200~$\AA$, 6900~$\AA$)     & 3-Gaussian & 2-Gaussian \\
\#2 58905 & LDT       & (4500~$\AA$, 5160~$\AA$)   & (6300~$\AA$, 6800~$\AA$)     & 3-Gaussian & 2-Gaussian \\ 
\#3 59105 & LDT       & (4500~$\AA$, 5160~$\AA$)   & (6300~$\AA$, 6800~$\AA$)     & 2-Gaussian & 2-Gaussian \\
\#4 59189 & LDT       & (4500~$\AA$, 5160~$\AA$)   & (6300~$\AA$, 6800~$\AA$)     & 2-Gaussian & 2-Gaussian \\
\#5 59468 & SAAO      & (4500~$\AA$, 5160~$\AA$)   & (6300~$\AA$, 6800~$\AA$)     & 2-Gaussian & 1-Gaussian \\
\#6 59586 & SALT      & (4500~$\AA$, 5160~$\AA$)   & (6300~$\AA$, 6760$~\AA$)     & 2-Gaussian & 1-Gaussian \\
\#7 59660 & FORS2     & (4500~$\AA$, 5160~$\AA$)   & (6300~$\AA$, 6800~$\AA$)     & 2-Gaussian & 2-Gaussian \\
\#8 59828 & DBSP      & (4500~$\AA$, 5160~$\AA$)   & (6300~$\AA$, 6800~$\AA$)     & 2-Gaussian & 1-Gaussian \\
\#9 59881 & SALT      & (4500~$\AA$, 5160~$\AA$)   & (6300~$\AA$, 6800~$\AA$)     & 2-Gaussian & 2-Gaussian \\
\#10 59883 & SALT     & (4500~$\AA$, 5160~$\AA$)   & (6300$\AA$, 6800$\AA$)       & 2-Gaussian & 1-Gaussian \\
\#11 60018 & FORS2    & (4500~$\AA$, 5160~$\AA$)   & (6300$\AA$, 6800$\AA$)       & 2-Gaussian & 2-Gaussian \\
\hline
\end{tabular}
\tablefoot{Columns 3 and 5 list the wavelength ranges used for local
fitting of the broad Balmer profiles line regions.
Columns 5 and 6 lists the (signal/noise-dependent) model used for fitting
the [\ion{O}{iii}]~$\lambda\lambda$4959,5007 and H$\beta$ profiles, respectively.}
\end{table*}

\begin{table*}[h!]
\centering
\caption{Best-fitting [\ion{O}{iii}]$\lambda$5007, H$\beta$ broad and narrow, [\ion{N}{ii}]$\lambda$6585, H$\alpha$ broad and narrow emission line fluxes, and $R_{\rm H\beta/[\ion{O}{iii}]}$}\label{tab:spec_comp2}
\begin{tabular}{lc|c|cccc|cc|c}
\hline
\hline
Spec. & Time    & spectral & \multicolumn{4}{c|}{Narrow line flux} & \multicolumn{2}{c|}{Broad line flux} & $R_{\rm H\beta/[\ion{O}{iii}]}$\\ 
     &   & index & [\ion{O}{III}]$\lambda$5007 & H$\beta$ & [\ion{N}{II}]$\lambda$6585 & H$\alpha$ & H$\beta$  & H$\alpha$\\
\hline
\#1 & 58871 & $-2.81 \pm 0.04$ & $7.8 \pm 0.4$  & $ 0.2 \pm 0.1$   & $4.1 \pm 0.1$ &$2.2 \pm 0.1$ & $40.6 \pm 1.1$ & $116.0 \pm 1.0$ & $5.3 \pm 0.3$\\
\#2 & 58905 & $-1.8 \pm 0.1$ & $7.0 \pm 0.8$   & $<0.1$  & $5.4 \pm 0.3$  &$4.9 \pm 0.5$   & $41.0 \pm 2.1$  & $124.3 \pm 2.8$   & $6.1 \pm 0.8$\\
\#3 & 59105 & $-0.5 \pm 0.1$ & $8.0 \pm 0.2$   & $<0.3$  & $2.9 \pm 0.7$  &$3.1 \pm 0.7$ & $19.4 \pm 2.2$ &  $90.4 \pm 2.2$   & $2.5 \pm 0.3$ \\
\#4 & 59189 & $-0.6 \pm 0.1$ & $ 7.2 \pm 0.5$  & $<0.2$  & $6.4 \pm 1.4$  &$5.7 \pm 0.2$ & $18.7 \pm 2.1$ &  $80.6 \pm 2.8$  & $2.6 \pm 0.3$\\
\#5 & 59468 &   --           & $ 8.3 \pm 0.6 $ & $<1.5$  & $8.6 \pm 0.3$  &$7.0 \pm 0.3$ & $8.6 \pm 2.8$ & $54.5 \pm 3.1$ & $1.1 \pm 0.4$\\
\#6 & 59586 &   --           & $ 6.3 \pm 0.5 $ & $<0.4$  & $6.8 \pm 1.6$  &$4.2 \pm 1.6$ & $17.7 \pm 2.6$ & $88.7 \pm 3.1$  & $2.8 \pm 0.4$ \\
\#7 & 59660 &   --           & $ 9.1 \pm 0.6 $ & $<0.2$  & $2.6 \pm 0.6$  &$3.14 \pm 0.6$& $20.0 \pm 1.5$ &  $96.4 \pm 3.5$ & $2.3 \pm 0.2$\\
\#8 & 59828 &   --           & $6.8 \pm 0.3$   & $<0.5$  & $3.0 \pm 0.4$    &$2.2 \pm 0.3$& $9.8 \pm 2.4$ & $39.2 \pm 1.2$  & $1.5 \pm 0.4$\\
\#9 & 59881 &   --           & $ 6.2 \pm 0.3 $ & $<0.1$  & $5.5 \pm 0.2$   &$4.0\pm 0.2$ & $7.0 \pm 1.5$ & $29.5 \pm 1.3$ & $1.0 \pm 0.2$\\
\#10 & 59883 &  --           & $ 5.4 \pm 0.5 $ & $<0.4$  & $5.4 \pm 1.4$    &$3.2 \pm 1.4$ & $9.0 \pm 3.2$ & $40.5 \pm 2.7$ & $1.6 \pm 0.6$\\
\#11 & 60018 &  --           & $8.0 \pm 0.6$   & $<0.1$  &$4.9 \pm 0.6$    &$3.15 \pm 0.6$ & $18.6 \pm 2.0$ & $76.4 \pm 1.7$ & $2.3 \pm 0.2$\\
\hline
\end{tabular}
\tablefoot{The broad line fluxes are estimated from the phenomenological Gaussian fitting. For the H$\alpha$ emission line, the flux value is the integrated flux for the best-fitting double Gaussian model. For the H$\beta$ emission line, most flux values are estimated from the double Gaussian fits, except those where we use a single Gaussian model due to low signal/noise. All flux units are $10^{-15}$~erg s$^{-1}$ cm$^{-2}$.} 
\end{table*}

\end{appendix}

\end{document}